\documentclass[
 reprint,
 amsmath,amssymb,
 aps,
]{revtex4-2}

\usepackage{graphicx}
\usepackage{dcolumn}
\usepackage{bm}
\usepackage{hyperref}
\hypersetup{colorlinks=true, citecolor=blue, urlcolor=blue, linkcolor=blue}
\usepackage{float}
\usepackage{algorithmicx}
\usepackage{algorithm}
\usepackage{algpseudocode}
\usepackage{color}
\usepackage{xcolor}
\usepackage[section]{placeins}

\begin{document}

\preprint{APS/123-QED}

\title{Prediction of social dilemmas in networked populations via graph neural networks}

\author{Huaiyu Tan$^1$}
\author{Yikang Lu$^1$}
\author{Alfonso de Miguel-Arribas$^2$}

\author{Lei Shi$^{1,3,}$}
\email{lshi@ynufe.edu.cn}

\affiliation{
1. School of Statistics and Mathematics, Yunnan University of Finance and Economics, 650221, Kunming, P. R. China\\
2. Zaragoza Logistics Center (ZLC), 50018, Zaragoza, Spain \\ 
3. Interdisciplinary Research Institute of data science, Shanghai Lixin University of Accounting and Finance, 201209, Shanghai, P. R. China
}

\date{\today}

\begin{abstract}
Human behavior presents significant challenges for data-driven approaches and machine learning, particularly in modeling the emergent and complex dynamics observed in social dilemmas. These challenges complicate the accurate prediction of strategic decision-making in structured populations, which is crucial for advancing our understanding of collective behavior. In this work, we introduce a novel approach to predicting high-dimensional collective behavior in structured populations engaged in social dilemmas. We propose a new feature extraction methodology, Topological Marginal Information Feature Extraction (TMIFE), which captures agent-level information over time. Leveraging TMIFE, we employ a graph neural network to encode networked dynamics and predict evolutionary outcomes under various social dilemma scenarios. Our approach is validated through numerical simulations and transfer learning, demonstrating its robustness and predictive accuracy. Furthermore, results from a Prisoner's Dilemma experiment involving human participants confirm that our method reliably predicts the macroscopic fraction of cooperation. These findings underscore the complexity of predicting high-dimensional behavior in structured populations and highlight the potential of graph-based machine learning techniques for this task.
\\ \\
\keywords: collective behavior prediction, social dilemmas, structured populations, graph neural networks.
\end{abstract}

\maketitle

\section{Introduction}
\label{sec:introduction}

As scientific efforts increasingly focus on addressing more complex problems and achieving more accurate predictions of phenomena, the need for new techniques becomes evident. In recent years, the availability of massive amounts of data, coupled with advances in computational power, has paved the way for a shift towards data-driven and statistical learning approaches \cite{kutz2016dynamic, brunton2021data, wang2021bridging, ghadami2022data, yu2024learning}. Machine learning and deep learning techniques are now being widely implemented across various fields. There is growing interest in leveraging these data-driven methods to tackle previously insurmountable challenges in the modeling and prediction of dynamical systems across numerous applications. These include, but are not limited to, health sciences \cite{hussein2018epileptic, solaija2018dynamic, woldaregay2019data, fu2020novel, kissas2020machine}, biology \cite{yeung2021data, eslami2022prediction}, epidemiology \cite{proctor2015discovering, bhouri2021covid}, ecology and climate science \cite{rasp2018deep, scher2018toward, christin2019applications, rammer2019harnessing, karevan2020transductive}, financial markets and economics \cite{mann2016dynamic, nosratabadi2020data, maliar2021deep}, and engineering \cite{mohan2018data, bhatnagar2019prediction, li2019deep, lu2019data, himanen2019data, wang2020towards, raissi2020hidden, brunton2020machine}, among many others.

Human behavior and strategic decision-making in social dilemmas is another paramount example of complex phenomena. Social dilemmas highlight the tension between individual self-interest and collective welfare, where personal gain can lead to poor outcomes for the group, as seen in issues like the evolution of cooperation \cite{axelrod1981evolution} and the management of shared resources \cite{hardin1968tragedy}. Historically, game theory \cite{von1953theory, nash1950equilibrium}, and later evolutionary game theory (EGT) \cite{smith1982evolution, weibull1997evolutionary, richter2023half}, have provided a conceptual and analytical framework to model strategic decision-making and social dilemma scenarios. In particular, EGT has been successful through the introduction of bounded rationality \cite{simon1990bounded, samuelson1995bounded}, the consideration of dynamic processes, and its extension to structured populations \cite{szabo2007evolutionary}. In recent decades, game theory and the study of human behavior in general have also benefited from data-driven and learning techniques, improving our modeling, prediction capabilities, and understanding of human rationality \cite{hartford2016deep, noti2016behavior, rosenfeld2018predicting, bourgin2019cognitive, peterson2021using}.

As we move toward a world with increased human-machine interactions, we are witnessing the expansive use of artificial intelligence across multiple real-life settings \cite{crandall2018cooperating, ishowo2019behavioural, mcintosh2023google}, such as autonomous vehicles \cite{bonnefon2016social, greene2016our}, morality-imbued agents assisting decision-making \cite{awad2018moral}, or generative agents acting as extensions of the human mind to stimulate new ideas and assist in tasks, like Large Language Models \cite{lu2024llms, ray2023chatgpt}, Stable Diffusion \cite{esser2024scaling}, and others. As a result, effectively identifying decision-making patterns in large groups of agents within critical settings, such as social dilemmas, has become a pressing need.

To date, EGT has primarily relied on analytical approaches, such as mean-field theories, critical phenomena-based tools, and Monte Carlo simulations \cite{perc2017statistical, traulsen2023future}, to characterize system behavior and the evolution of strategies in populations across various scenarios. In this work, we aim to contribute to the literature on data-driven and deep learning techniques in complex systems, particularly within evolutionary game theory, by focusing on the prediction of evolutionary social dilemmas in multi-agent networked systems via graph neural networks.

Some works have explored similar areas but with fundamental differences, none fully representing the system under study within the EGT framework. Nay et al. \cite{nay2016predicting} used a large number of human-subject experiments to build a model predicting human cooperation in repeated Prisoner's Dilemma (PD) games. Kolumbus and Noti \cite{kolumbus2019neural} predicted the actions of human players in repeated strategic interactions using multi-layer perceptrons and convolutional neural networks, though their focus was on $2 \times 2$ normal-form games. Vazifedan et al. \cite{vazifedan2022predicting} presented a deep convolutional neural network model for predicting human behavior in repeated games with size-variant payoff matrices. Lin et al. \cite{lin2022predicting} aimed to predict human decision-making sequences in different tasks, including iterated PD games, by using long short-term memory networks and leveraging human data from various studies. Despite the diversity of games and methodologies in these works, they all share a common focus on 1 vs. 1 scenarios, which contrasts with our approach of addressing networked multi-agent systems. To the best of our knowledge, this is the first work to use deep learning techniques to predict outcomes of social dilemma games in structured populations.

Our methodology consists of two stages.  First, we extract features by gathering topological marginal information (TMIFE), that is, relevant information for the dynamical process on the network, at the microscopic (agent-based) level. Second, we input the tensor of marginal information for each agent $N$ at $T$ time-slices, along with the adjacency matrix representing the agents' contact network, into graph convolutional network (GCN) that predicts each agent's strategy in the next time step.

Achieving high-dimensional prediction tasks, that is, involving a multi-agent system, can be very challenging due to the complexity of both behavioral evolutionary dynamics and network topology. We summarize the general challenges associated with these tasks. Fundamentally, the complexity of population strategy choices in spatial games stems from topological features, including heterogeneity, assortativity, and clustering. This complexity presents challenges for predicting spatial dynamical systems. The main obstacles include: (1) Unlike well-mixed populations, those defined on non-Euclidean spaces cannot be represented explicitly by replicator dynamic equations or exactly solved using mean-field and pairwise approximation theories. (2) Observing the behavior of the entire network and its components may be impractical or impossible due to privacy concerns or limitations in data collection, meaning external observations are often incomplete or marginal. (3) Components frequently exhibit high levels of nonlinearity and interdependence, which hinders the generalization of prediction problems over time scales and the inscription of higher-order information transfer on topologies through Markov processes.

Thus, to summarize, the major contribution of our work is pioneering the extension of multi-agent dynamical systems prediction tasks to higher dimensions. The high-dimensional behavioral prediction task we propose specifically refers to (1) coupling with evolutionary dynamics on a networked population, and (2) predicting spatial evolutionary patterns of group behavior in a social dilemma framework. Our work, based on multi-agent systems, predicts not only the dynamical trajectories of macroscopic observables but also the spatially structured evolutionary patterns of social dilemmas in structured populations. Furthermore, our method enables prediction using incomplete marginal data. By predicting collective behavior evolution under varying circumstances, our work offers a novel paradigm for understanding, analyzing, and simulating evolutionary social dynamics.

\section{Results}
\label{sec:results}

\begin{figure*}[htbp]
    \centering
    \includegraphics[width=0.90\linewidth]
    {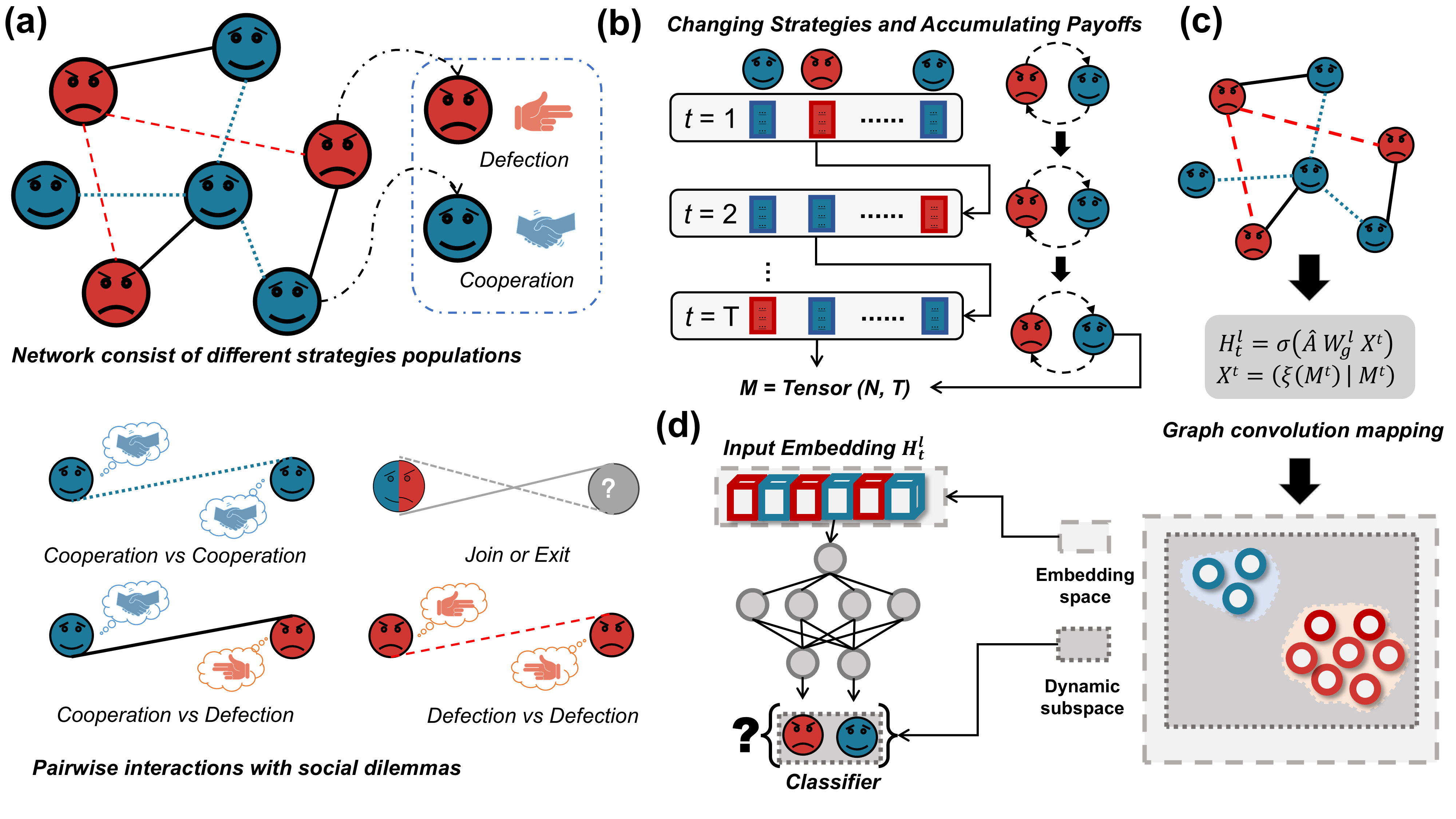}
    \caption{
        \textbf{TMIFE+GCN workflow scheme.}
        Different coloured nodes not only represent different strategies, but also indicate the higher-order features in the network. (a) A system of $N$ agents represented as a network with pairwise interactions. Agents have a certain property/attribute that may evolve in time under specific dynamical laws. (b) As the system evolves in time, marginal information is sampled from the $N$ agents at $T$ time slices. This information is encoded in the marginal data matrix $M$. For the case of game dynamics, this marginal information refers to the $N$ individual payoffs at each of the sampled $T$ slices. (c) Transformation of pre-encoded features into low-dimensional continuously dense embedding spaces using a graph neural network architecture. (d) Matching a classifier head to a graph embedding for performing a behavior prediction task.
    }
    \label{fig:fig_main_workflow}
\end{figure*}

Before presenting the main results, Fig. \ref{fig:fig_main_workflow} illustrates  the schematics of our Topological Marginal Information Feature Extraction + Graph Convolutional Networks (TMIFE+GCN) methodology. We start with a system of agents arranged as a network of pairwise interactions, where each agent is endowed with certain attributes or properties that may change over time according to specific dynamical laws (Fig. \ref{fig:fig_main_workflow}a). In this case, we focus on the dynamics of social dilemmas, where each agent is endorsed with a strategy used in pairwise interactions and a payoff that evolves based on the specific game interactions. From the dynamics exhibited by such a system, we collect marginal information for each of the $N$ agents at a certain number of time steps $T$, which includes individual payoffs (Fig. \ref{fig:fig_main_workflow}b). Next, the system's collective and topological information is encoded using the TMIFE method (Fig. \ref{fig:fig_main_workflow}c). Finally, the prediction of evolutionary dynamics is performed using a graph convolutional neural network architecture with a simple prediction head (Fig. \ref{fig:fig_main_workflow}d).

We consider two groups of evaluation metrics for each of the prediction tasks, the accuracy and F1-score metrics (details on \textbf{Supplementary Material}). These metrics have been widely used in a number of studies for performance on node state prediction tasks \cite{ref55, ref56, ref57}. 

To implement our methodology we utilize a well-known Python's neural network library, PyTorch \cite{ref54}. Specific parameter settings and numerical simulation scenario details are provided in the \textbf{Supplementary Material}. 

\subsection*{Learning the Prisoner's Dilemma game}
\label{subsec:learning_pdg}

\begin{figure*}[htbp]
    \centering
    \includegraphics[width=0.90\linewidth]
    {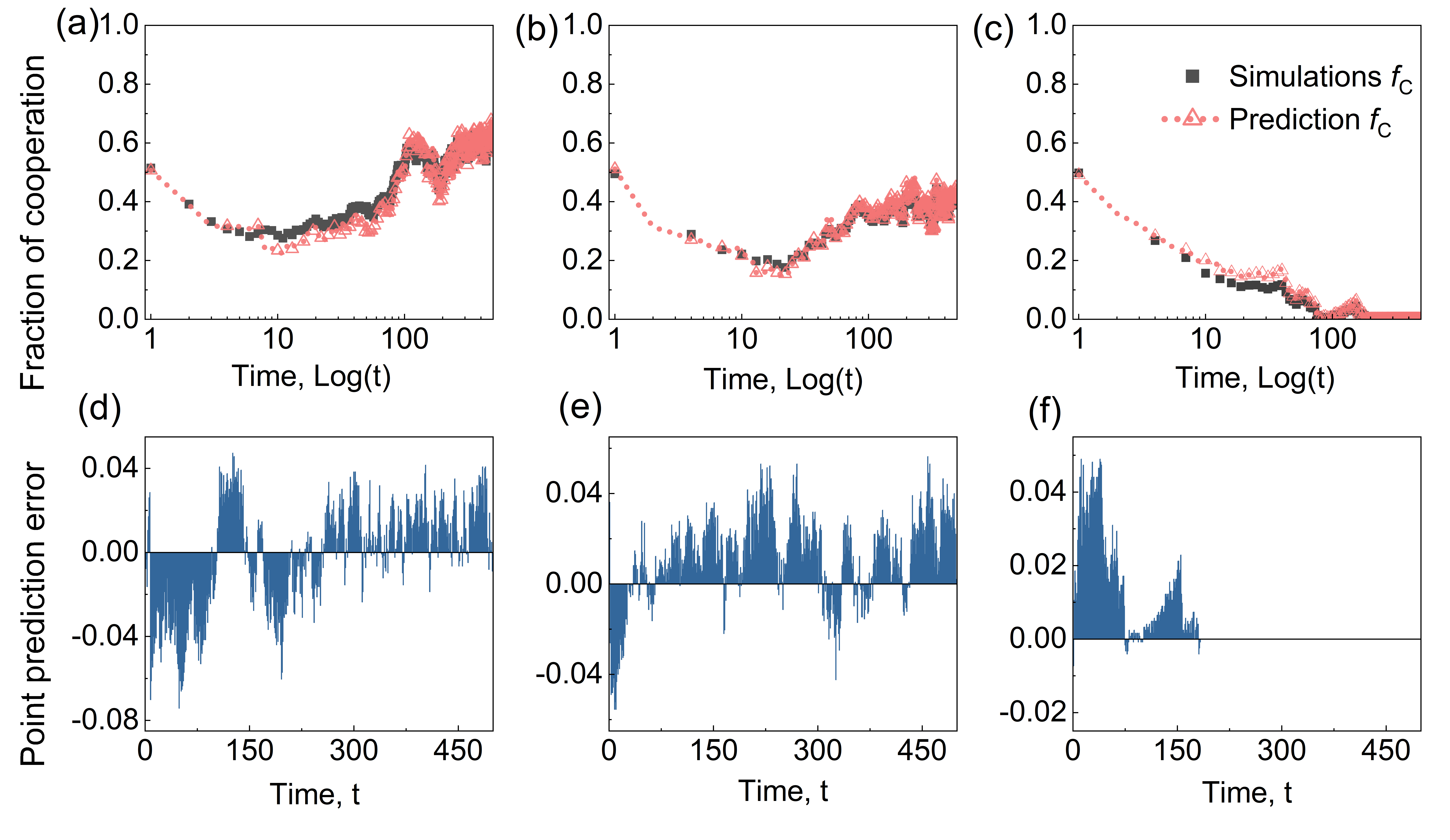}
    \caption{
        \textbf{Prediction of cooperative behavior in Prisoner's Dilemma.}
        Top panels represent the evolutionary time series of the fraction of cooperators from simulation results (black squares) and model prediction (red triangles), for different dilemma strengths $b$, $b=1.005$ (panel a), $b=1.015$ (panel b), and $b=1.035$ (panel c). Bottom panels represent the prediction's error for the same dilemma scenarios, respectively, in panel d to f.
    }
    \label{fig:fig_main_pdg_evolution}
\end{figure*}

As a first sample of the method's performance, we begin by testing it on the classical two-strategy Prisoner's Dilemma game (PDG) on a regular square lattice with length $L=50$, yielding a system size of $N=2500$ agents, under different temptation parameter $b$ (dilemma strengths) scenarios. Figure \ref{fig:fig_main_pdg_evolution} depicts the comparison of the evolutionary change of the fraction of cooperators in the system as obtained from the microscopic simulations and from the TMIFE+GCN methodology as well as the error between simulations and predictions, for $b=1.005$ (Panels a and d), $b=1.015$ (Panels b and e), and $b=1.035$ (Panels c and f). The mean square error (MSE) across time steps for each of these $b$ values is, respectively, $0.0076$, $0.0109$, and $0.0082$. Even though fluctuations are apparent due to finite-size effects, the prediction remains unaltered. As indicated by the point error, our method tends to underestimate for low $b$, and overestimate as $b$ grows.

We extend the prediction of the PDG to other topologies other than the square lattice (SL) network, such as the Barabási-Albert (BA) scale-free network and the Erdős-Rényi (ER) random network model. We found (see Table \ref{tab:table1}) that the prediction is better for dynamics on heterogeneous networks (BA) than in homogeneous ones (ER or SL). Accuracy is maintained at a high level as the average degree of the network increases. For BA networks, the testing set accuracy peaked at $98\%$. This indicates that our method is more sensitive to the density of the network, suggesting that for overly dense networks, the prediction performance becomes inaccurate due to the exponential increase in system complexity. Here, 'std.' represents the standard error calculated from repeating the experiment $100$ times to ensure that any noise in the results is minimized. In the same scenario, we predicted the evolutionary trajectory of the fraction of cooperation over time. The complete results under different network topology are provided in Table \ref{tab:table1}. The predictions of the evolutionary curves are also remarkably accurate, although a small fraction of moments exhibit some prediction bias (see Table \ref{tab:table2}). This outcome reflects a well-trained model generalized to a zero-shot dataset, which is certainly encouraging.

\begin{table}[htbp]
\centering
\caption{Test set evaluation for PDG. \label{tab:table1}}
\begin{tabular}{ccccccccc}
\hline
\multicolumn{1}{l}{}       & \multicolumn{1}{l}{Acc} & \multicolumn{1}{l}{F1} & \multicolumn{1}{l}{TPR} & \multicolumn{1}{l}{FPR}     & \multicolumn{1}{l}{Acc} & \multicolumn{1}{l}{F1} & \multicolumn{1}{l}{TPR} & \multicolumn{1}{l}{FPR} \\ \hline
$\langle k \rangle$ & \multicolumn{4}{c|}{BA}                                                                                  & \multicolumn{4}{c}{BA std.}                                                                          \\ \hline
2                          & 0.9748                  & 0.9815                 & 0.9828                  & \multicolumn{1}{c|}{0.3034} & 0.0473                  & 0.0444                 & 0.0554                  & 0.2771                  \\
4                          & 0.9548                  & 0.9721                 & 0.9543                  & \multicolumn{1}{c|}{0.0158} & 0.1067                  & 0.0692                 & 0.1083                  & 0.0805                  \\
6                          & 0.9087                  & 0.9499                 & 0.9187                  & \multicolumn{1}{c|}{0.4592} & 0.1469                  & 0.0937                 & 0.1469                  & 0.2919                  \\
8                          & 0.7217                  & 0.8281                 & 0.7217                  & \multicolumn{1}{c|}{0.6540} & 0.1504                  & 0.1126                 & 0.1504                  & 0.1743                  \\
10                         & 0.4993                  & 0.4878                 & 0.7840                  & \multicolumn{1}{c|}{0.7002} & 0.0556                  & 0.1573                 & 0.1251                  & 0.1509                  \\ \hline
                           & \multicolumn{4}{c|}{ER}                                                                                  & \multicolumn{4}{c}{ER std.}                                                                          \\ \hline
2                          & 0.8623                  & 0.9004                 & 0.9278                  & \multicolumn{1}{c|}{0.3581} & 0.0796                  & 0.0788                 & 0.1244                  & 0.0719                  \\
4                          & 0.8757                  & 0.9087                 & 0.9026                  & \multicolumn{1}{c|}{0.7783} & 0.1255                  & 0.1142                 & 0.1474                  & 0.2799                  \\
6                          & 0.8731                  & 0.9224                 & 0.8743                  & \multicolumn{1}{c|}{0.1348} & 0.1545                  & 0.0982                 & 0.1522                  & 0.2854                  \\
8                          & 0.7552                  & 0.8480                 & 0.7773                  & \multicolumn{1}{c|}{0.6899} & 0.1151                  & 0.0912                 & 0.1259                  & 0.1959                  \\
10                         & 0.7582                  & 0.8483                 & 0.7653                  & \multicolumn{1}{c|}{0.7013} & 0.1626                  & 0.1148                 & 0.1686                  & 0.2954                  \\ \hline
                           & \multicolumn{4}{c|}{SL}                                                                                  & \multicolumn{4}{c}{SL std.}                                                                          \\ \hline
4                          & 0.7607                  & 0.6854                 & 0.7875                  & \multicolumn{1}{c|}{0.2622} & 0.0543                  & 0.1275                 & 0.0724                  & 0.0620                  \\ \hline
\end{tabular}
\end{table}

\begin{table}[htbp]
\centering
\caption{Zero-shot evolutionary curve prediction for PDG. \label{tab:table2}}
\begin{tabular}{cccclll}
\hline
\multicolumn{1}{l}{}       & \multicolumn{1}{l}{MSE}    & \multicolumn{1}{l}{MAE}    & \multicolumn{1}{l}{JS}      & MSE     & MAE     & JS      \\ \hline
$\langle k \rangle$ & \multicolumn{3}{c|}{BA}                                                               & \multicolumn{3}{c}{BA std.} \\ \hline
2                          & 0.0004                     & 0.0090                     & \multicolumn{1}{c|}{0.0082} & 0.0011  & 0.0075  & 0.2670  \\
4                          & 0.0009                     & 0.0183                     & \multicolumn{1}{c|}{0.0143} & 0.0051  & 0.0073  & 0.2814  \\
6                          & 0.0033                     & 0.0352                     & \multicolumn{1}{c|}{0.0504} & 0.0095  & 0.0138  & 0.3047  \\
8                          & 0.0065                     & 0.0573                     & \multicolumn{1}{c|}{0.1450} & 0.0121  & 0.0190  & 0.3189  \\
10                         & 0.0072                     & 0.0269                     & \multicolumn{1}{c|}{0.0421} & 0.0177  & 0.0732  & 0.2666  \\ \hline
                           & \multicolumn{3}{c|}{ER}                                                               & \multicolumn{3}{c}{ER std.} \\ \hline
2                          & 0.0009                     & 0.0230                     & \multicolumn{1}{c|}{0.0183} & 0.0003  & 0.0093  & 0.0107  \\
4                          & 0.0014                     & 0.0345                     & \multicolumn{1}{c|}{0.0209} & 0.0007  & 0.0074  & 0.0308  \\
6                          & 0.0026                     & 0.0352                     & \multicolumn{1}{c|}{0.0603} & 0.0093  & 0.0038  & 0.0834  \\
8                          & 0.0029                     & 0.0304                     & \multicolumn{1}{c|}{0.0247} & 0.0468  & 0.0649  & 0.0801  \\
10                         & 0.0059                     & 0.0522                     & \multicolumn{1}{c|}{0.1508} & 0.0433  & 0.0506  & 0.1327  \\ \hline
                           & \multicolumn{3}{c|}{SL}                                                               & \multicolumn{3}{c}{SL std.} \\ \hline
4                          & \multicolumn{1}{l}{0.0016} & \multicolumn{1}{l}{0.0335} & \multicolumn{1}{l|}{0.0376} & 0.0007  & 0.0015  & 0.1517  \\ \hline
\end{tabular}
\end{table}

Before proceeding with other predictive tasks, we extensively compared the TMIFE-GCN model with several popular methods in graph machine learning used as baselines. These baseline methods include Node2Vec \cite{ref50}, GraphSAGE \cite{ref51}, GAT \cite{ref52}, and GIN \cite{ref53}. A brief description of these methods can be found in the \textbf{Supplementary Material}. It must be noted that the Node2Vec method is unable to directly utilize the attribute information of the nodes. To enable comparison with other methods, we used a modification of the biased walk strategy in random sampling, specifically by multiplying the node transfer probability by the strategy learning probability defined by the Fermi's updating rule in the evolutionary social dilemmas (Eq. S3 of the \textbf{Supplementary Material}). The results of the evolutionary behavior prediction, and the evolutionary trajectory prediction are provided in \textbf{Supplementary Material} (Table S3) and (Table S4), respectively.

 Until now, our focus has been on predicting macroscopic or aggregated behavior and its time evolution. Traditional behavioral prediction tasks have emphasized forecasting macroscopic features, which are inherently low-dimensional. However, these patterns emerge from complex interactions among the microscopic constituents—the players involved in the social dilemma—and there is not always a univocal correspondence between a macroscopic configuration and its underlying microscopic state. By concentrating solely on the macroscopic level, we risk overlooking crucial microscopic details of the process. To address this, we advance our approach to high-dimensional prediction, aiming to forecast the status of individual nodes at different time points. Accurately predicting these structured features is a key distinction of our work compared to other machine learning-based prediction methods.

\begin{figure}[htbp]
    \centering
    \includegraphics[width=1\linewidth]
    {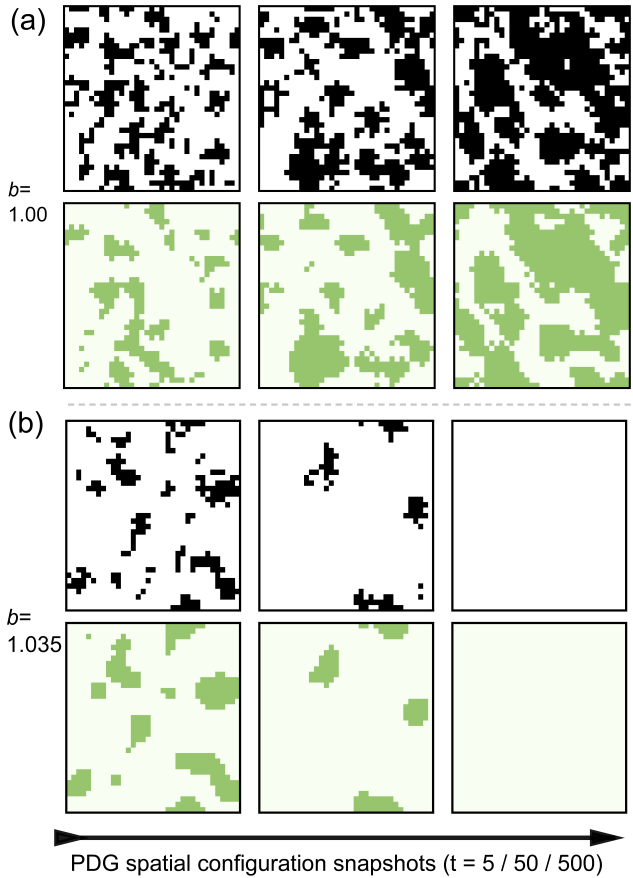}
    \caption{
        \textbf{PDG spatial configuration snapshots.} The method predicts the spatial evolutionary patterns of the PDG at both low and high dilemma strength levels. To facilitate observation of the self-organization phenomenon, time snapshots are taken at $t=5$, $50$, and $500$ on a $35 \times 35$ square lattice. Black and green colors indicate cooperation strategy from simulation and prediction, respectively. (a) Low dilemma strength ($b=1$), where cooperators form clusters to fend off defectors. (b) High dilemma strength ($b=1.035$), where cooperators initially form clusters, but due to the high levels of temptation, individuals eventually defect, leading to the extinction of cooperators. 
    }
    \label{fig:fig_main_pdg_spatial}
\end{figure}

Continuing with the PDG, we present this higher-dimensional prediction by examining the spatial patterns (configuration snapshots) of the agents' or nodes' strategies at specific time steps (Fig.~\ref{fig:fig_main_pdg_spatial}). We visually compare the differences between our PDG simulations and the TMIFE-GCN predictions at various time steps and dilemma strengths, as defined by the temptation to defect parameter $b$. The time snapshots illustrate the spatial pattern of strategy evolution on a $35 \times 35$ square lattice at $t=5$, $50$ and $500$ time steps. The smaller network size was chosen to balance computational complexity with pattern recognition effectiveness. Darker colors represent cooperation ($C$), while lighter colors represent defection ($D$).

\begin{figure}[htbp]
    \centering
    \includegraphics[width=0.9\linewidth]{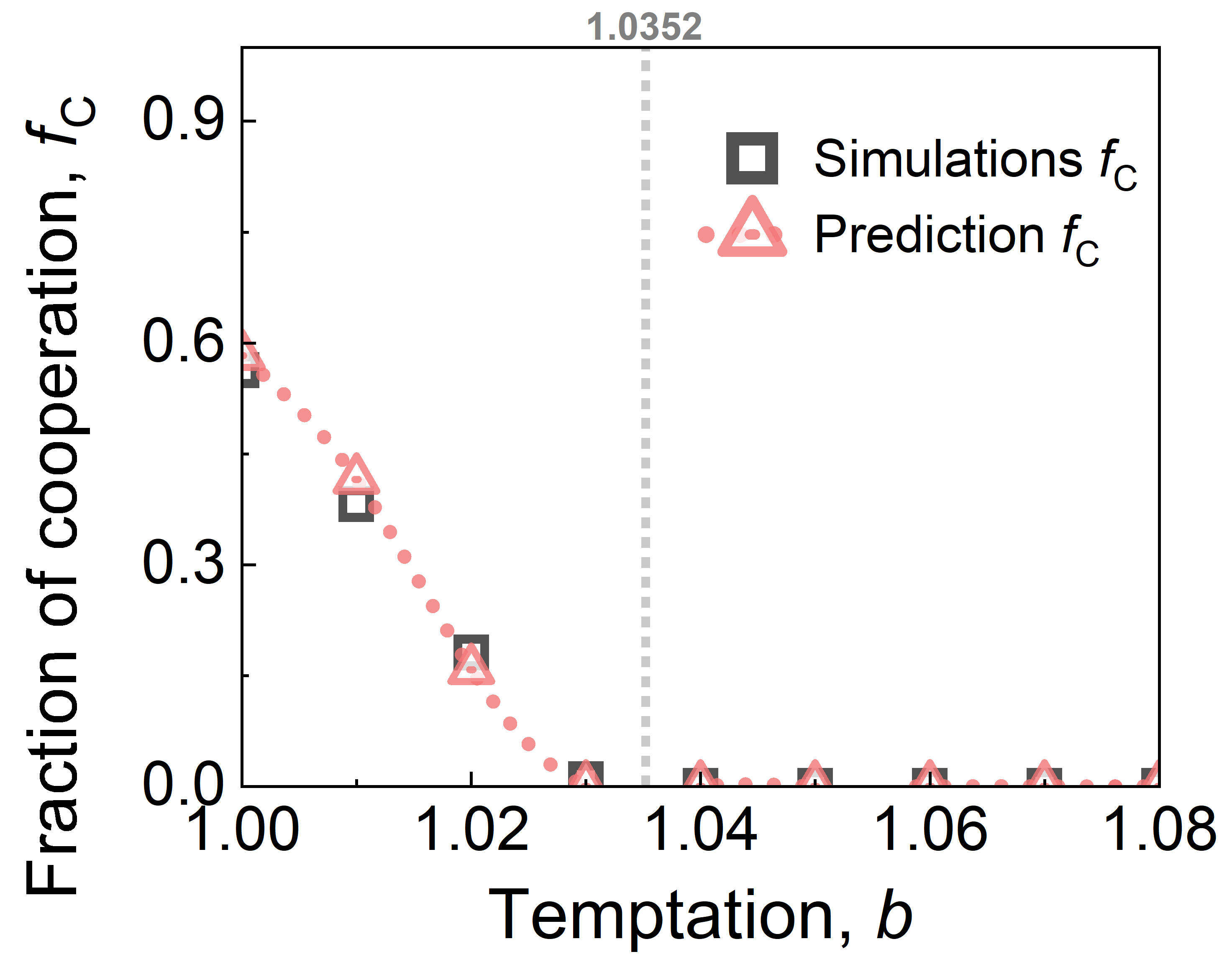}
    \caption{
        \textbf{Phase diagram in Prisoner's Dilemma.}
        The phase diagram represents the change of the fraction of cooperators $f_C$ against the change of the temptation to defect control parameter $b$. The diagram is generated both from simulations and from our prediction model. In this example, the case under study is that of a $50\times 50$ square lattice of agents with periodic boundary conditions. Black squares represent the simulation results and red triangles represent the prediction. The predicted critical point for cooperation extinction is found to be $b_c=1.035$.
        }
    \label{fig:fig_main_pdg_phase}
\end{figure}

Our method not only performs well in predicting evolutionary curves but also accurately reconstructs evolutionary trends, such as the clustering (or "hugging") of cooperators, as well as their survival and extinction under different dilemma strengths. Furthermore, the threshold for cooperation extinction is accurately predicted. We iteratively predict the full evolutionary steady state with respect to $b$ and identify the critical value of $b$ at which the phase transition towards cooperation extinction occurs, which is approximately $b_0=1.034$ \cite{ref58}. The prediction error is extremely small, demonstrating the effectiveness of the TMIFE-based approach. The prediction of the evolutionary phase transition is shown in Fig.~\ref{fig:fig_main_pdg_phase}.

These results suggest that our method can infer the evolutionary boundaries and thresholds of strategies without requiring large-scale computational simulations or human experiments. It only requires marginal data from a small fraction of time steps ($20\%$) to achieve good generalization ability. This provides a new perspective, based on graph convolution, for inferring the evolutionary trajectories of strategies and group behaviors in evolutionary games.

Finally, regarding the learning tasks of our model with respect the PDG, we study the effect of different population sizes and learning rates. We performed a robustness analysis and analyzed the graph embedding using t-SNE \cite{ref59} for dimensionality reduction. The robustness analysis was conducted on population size $N$ and learning rate $\alpha$ for SL, ER, and BA networks, respectively. The results are shown in \textbf{Supplementary Material} (Fig. S4). The graph embedding is represented in a $3$-dimensional space through t-SNE, referred to as the embedding space, which is shown in \textbf{Supplementary Material} (Fig. S5). The results indicate that with larger population sizes, the information is richer, and the prediction error becomes more stable. Additionally, a learning rate of $\alpha=0.06$ appears to be an optimal choice when training the model. In the visualized embedding space, our method offers two key advantages compared to using only adjacency matrices: (1) it captures more complete information, as seen in the comparison of the 3D visualizations in \textbf{Supplementary Material} (Fig. S5(a) vs. (b)), and (2) it provides clearer information boundaries, as seen in the comparison of the 2D visualizations in \textbf{Supplementary Material} (Fig. S5(c) vs. (d), and (e) vs. (f)).

\subsection*{Transfer learning from PDG}
\label{subsec:transfer}

\begin{figure*}[htbp]
    \centering
    \includegraphics[width=1\linewidth]{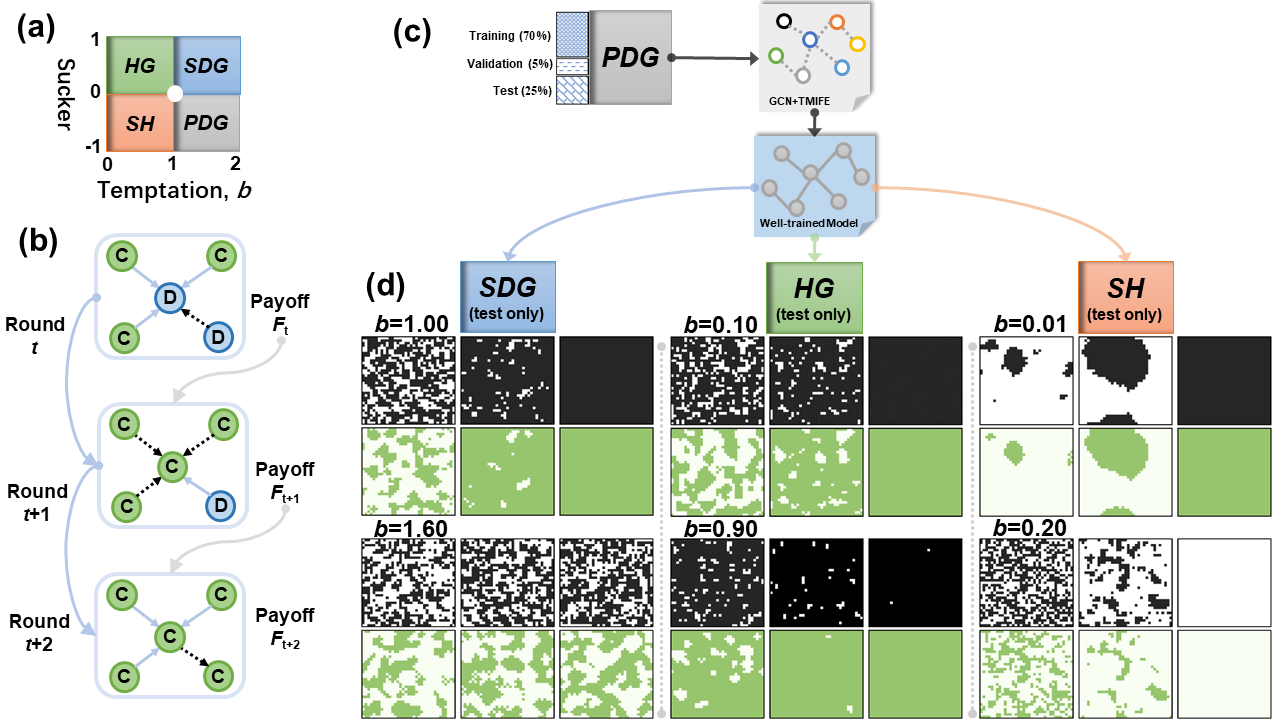}
    \caption{
        \textbf{Prediction of spatial patterns of collective behavior for transfer learning TMIFE+GCN method.} (a) Location of the Prisoner's Dilemma game (PDG), Snow-Drift game (SDG), and Harmony game (HG), and Stag-Hunt (SH) on the $S$-$b$ parameter space diagram. (b) Schematic diagram of the process by which cooperators form clusters during the evolutionary game dynamics. Green circles represent cooperation ($C$) and blue ones represent defection ($D$). The interaction neighbourhood of the game is indicated by arrows. The dotted line indicates that over time, neighbours in the neighbourhood influence the strategy choice for the next round based on the payoff $F_t$. (c) Basic logic of the transfer learning task. The TMIFE+GCN model is trained with PDG data, then is tested on different games other than the PDG. (d) Predictions of spatial pattern evolution under different dilemma strengths $b$ across different social dilemmas (SDG, HG, and SH).
        }
    \label{fig:fig_main_pdg_transfer_scheme}
\end{figure*}

\begin{figure}[htbp]
    \centering
    \includegraphics[width=1\linewidth]{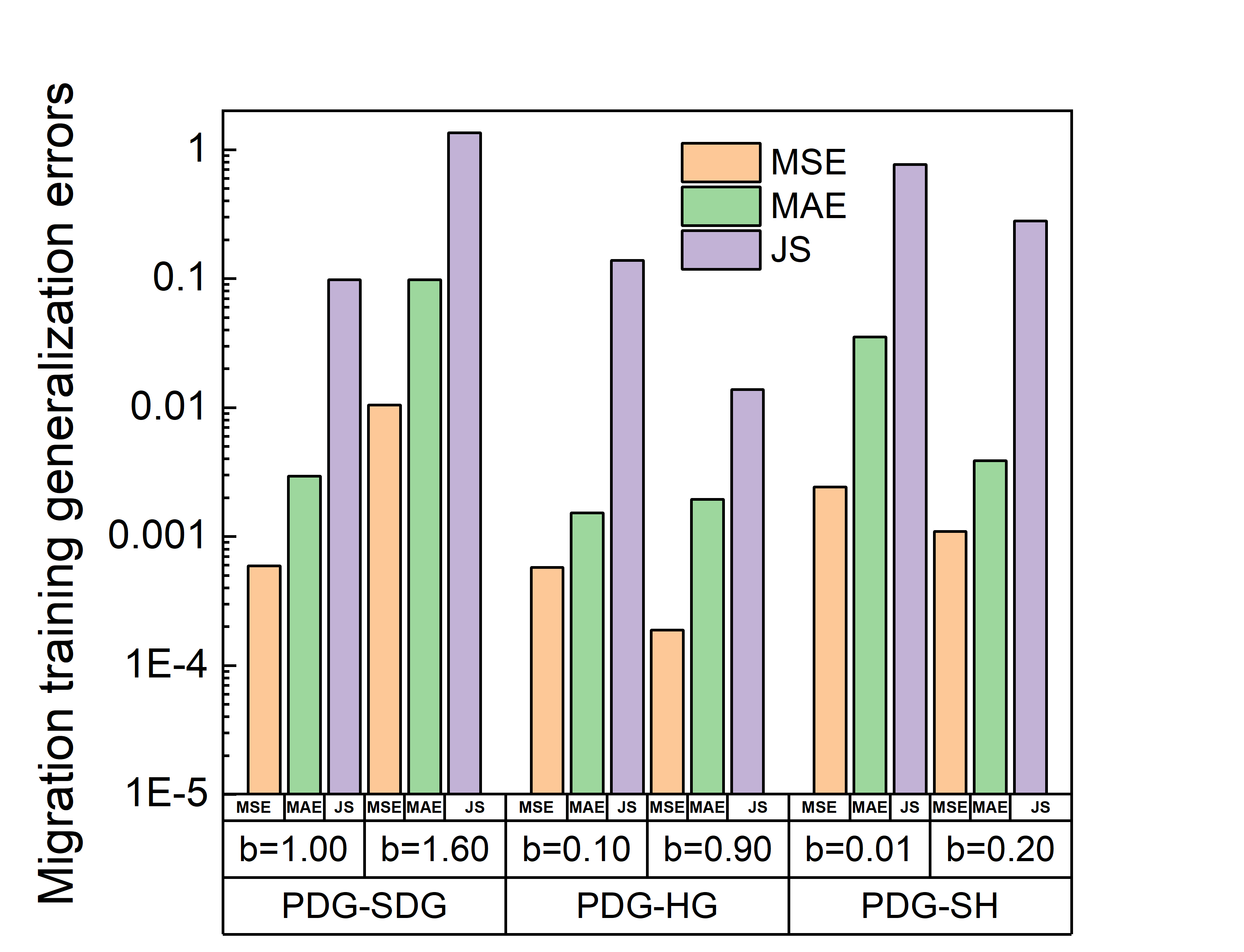}
    \caption{
        \textbf{Error in model's transfer learning.} Model transfer learning for multiple social dilemma scenarios (Snow Drift Game (SDG), Harmony Game (HG), and Stag-Hunt (SH)) was evaluated for a zero-shot behavior prediction task. The model's transfer learning performance is assessed using three metrics: mean square error (MSE), mean absolute error (MAE), and Jensen-Shannon divergence (JS). Each transfer task was performed once for two levels of dilemma strength, denoted by $b$.
        }
    \label{fig:fig_main_pdg_transfer_error.png}
\end{figure}

Now we ask about the generalization capabilities of our learning framework when facing prediction of related but genuinely different dynamical systems without previous exposure.  Thus, we tested the transfer learning performance of our methods to confirm that the prediction method truly learned group decision-making information from social dilemma scenarios. We conduct experiments under four social dilemmas (Fig.~\ref{fig:fig_main_pdg_transfer_scheme}(a)), and Fig.~\ref{fig:fig_main_pdg_transfer_scheme}(c) demonstrates the basic idea of the transfer learning process. In this case, we train the model on PDG, while zero-shot predicts the group behavior evolution on SDG, HG, and SH. The Temptation parameters were chosen for each game framework as SDG: $b=1.00, 1.60$; HG: $b=0.10, 0.90$; SH: $0.01, 0.20$. The model is training on PDG with $b=1.02$, and this parameter was tried by repeated attempts to get the optimal effect. The results of the transfer tasks are shown in (Fig.~\ref{fig:fig_main_pdg_transfer_scheme}(d)). Surprisingly, our methods is able to develop memorability for human behavioral patterns by learning on the PDG, and achieves good generalization capabilities in the SDG, HG, and SH game frameworks, even though the methods never experiences these additional social dilemma situations. However, the results also reflect the worse side, for example, the model trained on PDG learned the mapping of the rule that cooperators may form clusters to defend themselves against defectors, while overemphasizing this detail in other scenarios led to bad predictions. This phenomenon is most evident at (SDG, $b=1.60$). In addition, Fig.~\ref{fig:fig_main_pdg_transfer_scheme}(c) presents a schematic diagram of a cooperator resisting the invasion of a defector by forming clusters. Fig~\ref{fig:fig_main_pdg_transfer_error.png} presents the various accuracy metrics on the transfer learning tasks.

\subsection*{Human-played PDG}
\label{subsec:human}

\begin{figure}[htbp]
    \centering
    \includegraphics[width=1\linewidth]{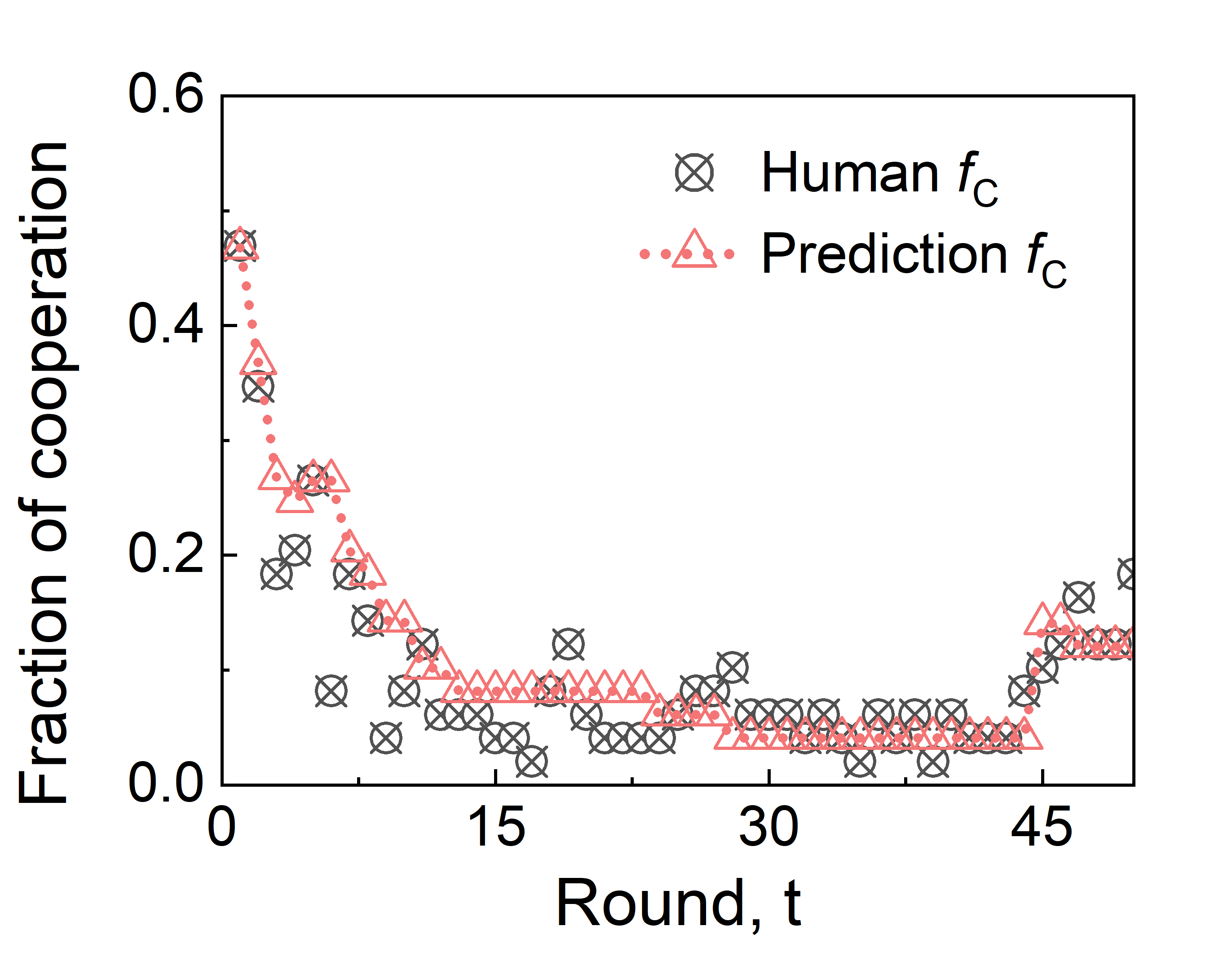}
    \caption{
        \textbf{Human-played PDG game experiment.} The black circles represent the observed human cooperation fraction, and the red triangles represent the learned cooperation fraction. The experiment is performed on a $7\times 7$ square lattice network with periodic boundary conditions. A total of $49$ individuals participated, and the iteration of the dynamics continued for $50$ rounds. 
        }
    \label{fig:fig_main_pdg_human}
\end{figure}

To demonstrate the practical capabilities of our proposed method, we also tested it in a real human behavioral experimental setting. Volunteers were recruited to organize a human-played PDG; detailed setting is provided in the \textbf{Supplementary Material}. The model was trained using only the first $5$ rounds of labeled data with real strategies and then used to make predictions at all time steps. As shown in Fig.~\ref{fig:fig_main_pdg_human}, our method accurately predicts the evolutionary change in the fraction of cooperators in the population. However, due to the small population size ($7 \times 7$ sites), the predictions exhibit some fluctuations.

\subsection*{Prediction for PDG with 'exit' strategy}
\label{subsec:three_strategy}

Finally, regarding evolutionary game dynamics, we investigate our models' prediction capabilities in a three-strategy extension of the PDG \cite{ref37}. This version of the PDG includes a third strategy referred to as 'exit' ($E$). Exiters, those who choose the exit strategy, receive a small, guaranteed payoff $\epsilon>0$, avoiding exploitation by defectors. This exit mechanism, thus, empowers players to temporarily leave the game and potentially face abusive interactions from other players. It must be noted that introducing a third strategy in the dynamics further complicates the high-dimensional prediction task as anticipated by the rich dynamical behavior found in \cite{ref37}. Correctly predicting the dynamics under exit strategy may indicate that our approach has the ability to learn decision-making patterns, rather than being a result of pure chance. 

\begin{figure*}[htbp]
    \centering
    \includegraphics[width=0.9\linewidth]{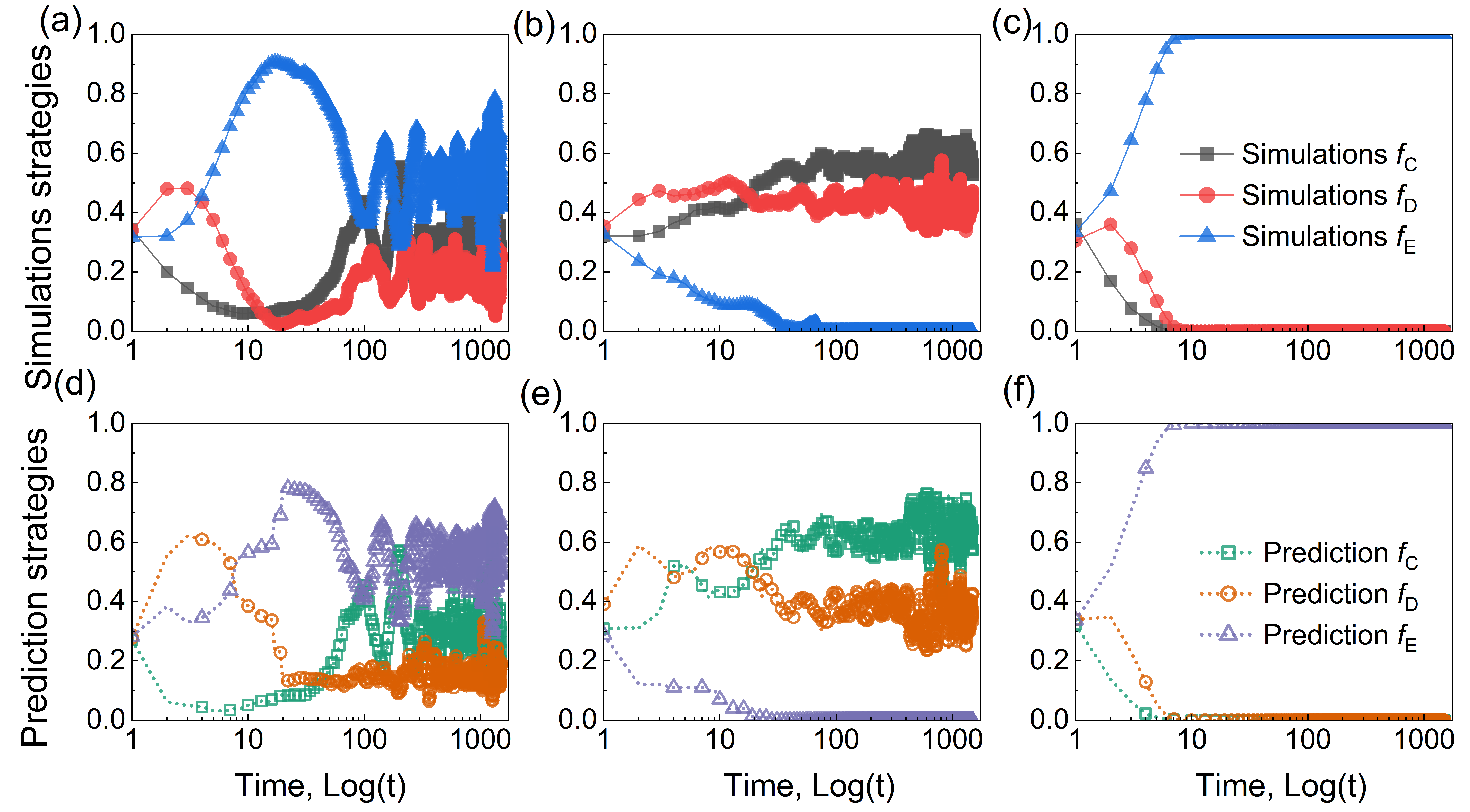}
    \caption{
        \textbf{Three-strategy extended PDG with 'exit' strategy.} Top panels (a to c) represent simulation results and bottom panels (d to f) represent the prediction produced by the TMIFE+GCN method. The y-axis depicts the time series of fraction of the three strategies (cooperation $C$, defection $D$, and exit $E$) under the evolutionary dynamics for $b=1.4$ and $\epsilon=0.2$ (Panels a and d), for $b=1.0$, $\epsilon=0.1$ (Panels b and e), and for $b=1.4$, $\epsilon=0.6$ (Panels c and f). Qualitatively, we can appreciate how the prediction matched the simulation results for the three different phases. Note that in the bottom row, for visualization purposes and software performance, only one data point is shown every two time steps.
        }
    \label{fig:fig_main_pdg_exit}
\end{figure*}

We explore this model for a set of $(b,\epsilon)$ control parameter values that give rise to three qualitatively different macroscopic phases. Coexistence of $C$ and $D$ (i.e. suppression of strategy $E$) for $b=1.4$, $\epsilon=0.2$; coexistence of the three strategies ($C,D,E$) for $b=1.0$, $\epsilon=0.1$, and absolute dominance of $E$ strategy (extinction of $C$ and $D$) for $b=1.4$, $\epsilon=0.6$. Figure \ref{fig:fig_main_pdg_exit} depicts the evolutionary time series of the fraction of strategies (top panels for the simulation results) and (bottom panels for the learning method's prediction). The increased number of strategic choices available increases the complexity of the dynamics and this might complicate prediction with respect to the classical two-strategy PDG. However, we can appreciate that our methodology correctly predicts the qualitatively different phases of the system. Moreover, our predictions demonstrate two key advantages: (1) better prediction of the cyclic dominance phenomenon in the three-strategy model, and (2) smaller differences between the predicted and true values for different parameters, even when inevitable evolutionary fluctuations occur.

\subsection*{Epidemic dynamics}
\label{subsec:epidemic}

Finally, we were curious to see if our method could be applied to learning and predicting different types of dynamics. We applied our methodology to a standard example of binary-state dynamics out of the evolutionary game area: the susceptible-infected-susceptible (SIS) dynamics for the spread of infectious diseases.  Specifically, we used a discrete-time SIS dynamics with recovery rate $\gamma=0.2$ and transmission rate $\beta=0.01$, $0.3$, and $0.7$, on a regular square lattice. At odds with the evolutionary game learning tasks, where the marginal data for the TMIFE method was the individual payoffs at every round, we now use the individual probability of being under infected health status at every time step, $P_i(X(t)=1)$, $i=1,...,N$, where $1$ represents the individual infected health status. The \textbf{Supplementary Material} provides the SIS model details, the test set accuracy and predictions (Tables S5 and S6), and the simulation-prediction comparison for the different $\beta$ values used (Fig. S3). Again, we observe a very good qualitative and quantitative agreement between the ground truth (simulations) and the prediction (TMIFE+GCN learning task).

\section{Discussion} 
\label{sec:discussion}

Our work addresses the high-dimensional prediction of collective behavior in structured populations under social dilemma scenarios by introducing a novel encoding method, TMIFE, for social dynamics features derived from limited marginal data. By integrating graph neural network algorithms, the model effectively captures both dynamic and topological features simultaneously.

We predict the evolutionary time series of agents' strategies across various topologies in a Prisoner's Dilemma game (PDG), the phase transition threshold for the extinction of cooperation, and the spatial organization of agents' strategies in a regular lattice. Notably, our model accurately predicts high-dimensional spatiotemporal dynamics using only the initial $20\%$ of training steps, without employing recurrent neural network architectures. We compared the model's performance with various baseline methods and conducted ablation experiments. Additionally, to evaluate the capabilities of our framework, we extended the training and prediction tasks to a three-strategy extension of the PDG while maintaining prediction accuracy. We also applied our framework to an epidemic Susceptible-Infected-Susceptible (SIS) model, where the underlying dynamics and interaction mechanisms differ from those of social games, achieving similarly strong results. Finally, we performed zero-shot predictions in diverse social dilemmas (Snow Drift, Stag-Hunt, and Harmony games) after training on the PDG.

To the best of our knowledge, this study is the first to address high-dimensional prediction of collective behavior in the context of social dilemmas. It advances the field of human behavior prediction by accurately forecasting spatial evolution patterns in complex network arrangements of individuals. Our findings have significant implications for understanding the emergence and evolution of collective behavior within social networks. Additionally, our work explores new possibilities for designing generative swarm agents, as our model can more precisely simulate group coordination and evolutionary strategy selection in social dilemmas. Considering the trade-off between efficiency and effectiveness, we refrained from studying larger-scale networks and temporal networks due to their complexity. Future work will address topics such as (1) generalizing the model to a larger scale, and (2) incorporating network topology evolution into feature extraction by integrating its dynamic processes.

\section{Methods} 
\label{sec:methods}

\subsection{Dynamical processes on networks} \label{subsec:dynamical}

Our TMIFE-GCN method is designed for predicting dynamical systems. We primarily focus on evolutionary game dynamics on networks, with a main emphasis on the Prisoner's Dilemma game (PDG). Several network topologies are considered, including a square lattice (SL), the Barabási-Albert (BA) scale-free network model, and the Erdős-Rényi (ER) random network model. To evaluate the performance of our method across different dynamical systems, we also consider a three-strategy PDG model \cite{ref37} that includes an 'exit' strategy, typical variations of the PDG such as the Snow-Drift, Stag-Hunt, and Harmony games, as well as the susceptible-infected-susceptible (SIS) model for epidemic spreading. Details of all these models, including specific parameter settings for both the networks and the dynamics utilized in the simulations, are provided in the \textbf{Supplementary Material}.

\subsection{Marginal Data and Multi-Interaction Encoding Framework} \label{subsec:marginal}

In practice, it is often impossible to observe dynamical systems in complete detail, resulting in limited access to comprehensive data. To elaborate on marginal data, intermediate variables are typically generated in real time on a per-individual basis due to internal interactions within the topology. Examples include the individual infection probability in epidemic systems or the pair-based payoffs in evolutionary game systems. However, obtaining fine-grained observations between these individual pairs can be difficult or costly. It is, therefore, practical to record information about each individual's interactions over time, such as payoffs or disease incidence. This summarized information, which disregards specific interactions, is referred to as marginal data. Our aim is to efficiently encode marginal data to fully exploit both behavioral and topological dynamics using graph neural networks and fully connected neural networks. The prediction task is essentially to classify the states of interacting individuals in a topological system and generalize across different time scales. The primary challenges involve (1) accurately predicting the phase transition of the system, (2) predicting spatial patterns of behavioral evolution, and (3) ensuring the generalization ability of the behavior prediction model. The proposed methods address problems (1) and (2) and partially solve problem (3). Details on obtaining marginal data under different dynamics and encoding interactions through networks can be found in the \textbf{Supplementary Material}.

\subsection{High-Dimensional Collective Behavior Prediction Based on Graph Convolutional Networks} \label{subsec:tmife}

Similar to black-box systems, topological systems can exhibit complex phenomena through simple rules. The key difference is that a topological system internally clarifies the connections between components, which creates challenges for dynamic prediction since modeling with specific equations or encoding information passing between links is often impractical. Graph representation learning enables the filtering of available topological features, and the message-passing mechanism is naturally suited for extracting link information. We propose a new feature encoding method, TMIFE, that performs well in graph convolution via a message-passing mechanism, requiring only minimal marginal data to efficiently encode inner interactions across various dynamic scenarios. We use graph convolution networks as the base model and explain the model's implementation of convolution via Chebyshev polynomial approximation \cite{ref43}, based on spectral graph theory \cite{ref39}. For specific model details and evaluation metrics, as well as supplementary results, see the \textbf{Supplementary Material}.

\subsection*{Code availability}
You can obtain the code by contacting the author.

\subsection*{Data availability}
You can obtain the Data by contacting the author.

\begin{acknowledgments}
L. Shi was supported by the key projects of the National Natural Science Foundation of China (NNSFC) (Nos. 11931015, 11671348 and 12271471). H. Tan was supported by the Scientific Research Foundation of Yunnan Education Department, China project no.2025Y, respectively. Also, thanks to Chen Shen from Kyushu University for providing experimental data and insights into the topic of evolutionary games.
\end{acknowledgments}

\subsection*{Author contributions}
Huaiyu Tan performed the method design and computational simulations. Yikang Lu and Alfonso de Miguel-Arribas wrote the manuscript, Alfonso provided insights into the behavioural dynamics part of the paper. Lei Shi organised real human-played social dilemma experiments and collected data.

\newpage

\bibliography{main_references}

\newpage

\end{document}


\maketitle

\section{Methods details}
\label{sec:methods}

\subsection{Evolutionary Game Dynamics}
\label{subsec:games}

Evolutionary games are powerful tools for studying coordination and cooperation behaviors in multi-agent systems. Through imitation-based pairwise interactions, we can explore how cooperation emerges and is sustained among rational agents. These games are defined by a payoff matrix, representing the gains or losses that players experience depending on the strategies employed by the players involved in the interaction. When the game process is influenced by a network structure, various complex phenomena can arise.

This network structure, or topology, defined by $\mathcal{G}=(\mathcal{V},\mathcal{E})$, consists of set of nodes $\mathcal{G}$ that represent the players in the game, while the links between nodes  (from the set $\mathcal{E}$) signify social ties or potential interactions. Players are assigned a strategy $s$, with cooperation ($C$) and defection ($D$) being the two archetypical options, that is, $s=\{C,D\}$.

If both the focal player and their opponent choose cooperation ($C$), they each receive a payoff of $R$. If the focal player chooses cooperation ($C$) while the opponent defects ($D$), the focal player receives a lower payoff $S$, and the opponent gains a higher payoff of $T$, representing the temptation to defect. If both players choose defection ($D$), they each receive a punishment payoff $P$.

The pairwise Prisoner's Dilemma Game (PDG) on a structured population can be represented by the following payoff matrix $M$:
\begin{equation}
\label{eq:payoff_matrix}
\begin{array}{l}
 \;\;\;\;\;\;\;\;\;\;\;
 \begin{array}{*{20}c}
   {C} & {D}  \\
 \end{array} \\ 
 \begin{array}{*{20}c}
   C  \\
   D  \\
\end{array}\;\left( {
\begin{array}{*{20}c}
   R & S  \\
   T & P \\
\end{array}
} \right),   \\ 
\end{array}
\end{equation}

It must be noted that these payoffs refer to instant payoffs yielded after a single round of the game. Depending on the specific relationship between the payoffs $R$, $S$, $T$ and $P$, different games can be defined:
\begin{itemize}
    \item{\textbf{Prisoner's Dilemma Game (PDG)}}, $R=1, S=0, T=b, P=0$, and $b \in [1, 2)$,
    \item{\textbf{Snow-Drift Game (SDG)}}, $R=1, S=0, T=b, P=-1$, and $b \in [1, 2)$,
    \item{\textbf{Harmony Game (HG)}}, $R=1, S=0, T=b, P=-1$, and $b \in [0, 1)$,
    \item{\textbf{Stag-Hunt (SH)}}, $R=1, S=-1, T=b, P=0$, and $b \in [0, 1)$.
\end{itemize}
Here, $b$ represents the temptation to defect parameter, which measures the strength of the dilemma. The greater the value of $b$, the higher the payoff for a defector exploiting a cooperator. Re-parameterizing $b$ allows for a simple and efficient way to control different social dilemma scenarios and their respective dilemma strengths. By adjusting the two parameters — Sucker's payoff ($S$) and Temptation ($T$) — we can switch between different games, highlighting the interrelationships among these games. 

Independently of the game under study, the evolutionary game dynamics proceed as follows. First, strategies are randomly assigned across the network of players. Then, the dynamic process begins and continues over iterative rounds. In each round, players interact with their immediate neighbors as defined by the network topology. This interaction consists of playing the specific game according to the strategies of the players involved. Instant payoffs are assigned to each player after each interaction, accumulating as every player plays against all their neighbors. This results in a cumulative payoff, which for a focal player $i$, is computed as:

\begin{subequations}
    \label{eq:cumulative_payoff}
    \begin{align}
        \Phi_i &= \sum_{j \in \Omega_i}{s_i^{\prime} M s_j}, \quad (i, j) \in \mathcal{E}, \label{eq:2A}\\
        s_i &=
        \begin{cases}
            (0, 1),\quad C\\
            (1, 0),\quad D
        \end{cases}, \quad i \in \mathcal{V}, \label{eq:2B}
    \end{align}
\end{subequations}

where $\Omega_i$ is the set of opponents directly connected to $i$ in $\mathcal{G}$, and $s$ is the set of strategies, represented as a column vector. Next, the strategy imitation is defined by the standard Fermi updating rule:

\begin{equation} 
\label{eq:fermi_updating} 
w(i \leftarrow j) = \dfrac{1}{1+\exp\{(\Phi_i-\Phi_j)/\kappa\}} 
\end{equation}

Here, $w(i \leftarrow j)$ indicates the probability that player $i$ will adopt the strategy of player $j$. If this probability is satisfied, player $i$'s strategy will be updated in the next round of the game. The parameter $\kappa$ represents the system's noise, reflecting the level of error, with respect to rationality, when facing the decision-making.

The macroscopic behavior of the system is characterized by the fraction of players adopting each of the available strategies, $f_s$, where $s = C, D$.

Additionally, in this work, we consider an extended version of the PDG that includes a third strategy, 'exit' ($E$) \cite{shen2021exit}. Alongside cooperation ($C$) and defection ($D$), players can now choose the exit strategy ($E$), which results in temporarily quitting the game for a stable but small payoff, $\epsilon$. The payoff matrix for this three-strategy extended PDG is:

\begin{equation}
\label{eq4}
\begin{array}{l}
 \;\;\;\;\;\;\;\;\;\;\;
 \begin{array}{*{20}c}
   {C} & {D} & {E} \\
 \end{array} \\ 
 \begin{array}{*{20}c}
   C  \\
   D  \\
   E  \\
\end{array}\;\left( {
\begin{array}{*{20}c}
   1 & 0 & 0 \\
   b & 0 & 0 \\
   \epsilon & \epsilon & \epsilon\\
\end{array}
} \right),   \\ 
\end{array}
\end{equation}

The dynamics of the game with three strategies proceed in the same way as for the two-strategy games.

In the first stage of this work, we train, test, and evaluate our model exclusively on the Prisoner's Dilemma Game (PDG). Subsequently, we assess the model's performance on other games without any prior training or exposure (a form of transfer learning). Later, we extend this approach by training and testing the model on both the three-strategy PDG and a human-played PDG (see details of this settings below). The overarching goal is to determine whether the model can learn the spatial evolutionary dynamics using only payoff margin data and topological information.

\subsection{Epidemic Dynamics}
\label{subsec:epidemics}

To investigate whether our model can accurately predict individual statuses in other dynamical processes, we also applied it to one of the standard dynamical processes in the literature: the susceptible-infected-susceptible (SIS) model for epidemic spreading of infectious diseases . Modeling epidemic spread through a network topology may give rise to a number of complex phenomena. Networks influence the trajectory of propagation and introduce complexity to dynamics through homogeneous or heterogeneous structures. 

For simplicity here, we primarily consider a square lattice (SL) topology. The SIS model consists of two stochastic processes: the contagion and the recovery processes. The probability of a contagion taking place assuming the focal node $i$ is susceptible (denoted by $S$ or $0$, here), and the probability that a recovery occurs given that the focal node $i$ is infected ($I$ or $1$), during time interval $\Delta t$, are respectively given by:
\begin{subequations} \label{eq:SIS}
\begin{align}
P(X_i(t+\Delta t)=1|X_i(t)=0) &= 1-(1-\beta\Delta t)^{z_i}, \label{eq:5A}\\
P(X_i(t+\Delta t)=0|X_i(t)=1) &= \gamma\Delta t, \label{eq:5B}
\end{align}
\end{subequations}
with $\Delta t \equiv 1$, and $z_i$ equal to the number of infected neighbors of focal node $i$, computed as: 
\begin{align} 
    \label{eq6} 
    z_i &= \sum_{j \in \Omega_i}{\text{sign}(X_j(t)=1)}, \quad e(i, j) \in \mathcal{E},
\end{align}
where $\text{sign}(\cdot)$ is an indicator function that takes the value $1$ when the condition is satisfied, and $\Omega_i$ is the set of neighbors of node $i$. At each time step, the agents statuses change according to the processes described above, continuing until the system reaches a stable state where the macroscopic fraction of infected individuals $f_I$, also known as the disease's prevalence, no longer changes.

To characterize this system's behavior, we focus on this fraction of infected individuals $f_I(t)$ at any given time step. For the SIS prediction task, the marginal data required refers to the individual probability of infection, $P(X_i(t+1)=1|X_i(t)=0)$, computed using Eq.\ref{eq:SIS}, under the assumption that specific propagation links are difficult to observe. Additionally, because $P(X_i(t)=1)$ cannot be calculated for susceptible individuals, the best approach is to assume $P(X_i(t)=1)=0$. In this task, the label is naturally whether an agent is infected or not ($I$, $S$), and our goal is to predict the system's steady state at any arbitrary time step, using limited marginal data on the dynamic processes. Similar to the PDG case, this is essentially an extension of the node classification task.

\subsection{Topology-based Marginal Interaction Feature Encoding (TMIFE)}
\label{subsec:tmife}

Graph Convolutional Networks (GCNs) can utilize attribute information associated with the nodes in a graph topology. However, when internal interactions within the system are present, effectively predicting downstream tasks becomes challenging due to difficulties in observing fine-grained attribute information. To address the problem of inadequately representing internal interaction dynamics in cases of incomplete data, we propose a novel feature encoding method: Topology-based Marginal Interaction Feature Encoding (TMIFE). TMIFE enables us to capture the internal interaction processes across the entire system using only marginal data. This method incorporates both global state information and local interaction details.

By relying on the message-passing framework of GCNs, TMIFE allows us to accurately infer node types and the system's evolution over time. Explainability analysis demonstrates that, compared to approaches relying solely on network structure, TMIFE produces sharper, smoother, and more stable node embeddings, leading to improved performance in downstream classification tasks.

Inner-interaction encoding with marginal data essentially involves finding a generalized function $\xi(\cdot)$ that maps the marginal data to interaction information. This mapping requires exploiting the message gap between pairs of nodes in the topological system. Let the marginal data at time $t$ be $\mathcal{M}_{t} = (m_1,..., m_N)^{\prime}$ on a complex network of size $N$, where the ${\prime}$ symbol here denotes a column vector. As the system evolves over $T$ steps, all marginal data are defined as $\mathcal{M} \in \mathcal{R}^{N \times T}$:

\begin{equation}
\label{eq:marginal_data}
\mathcal{M}=
\begin{pmatrix}
 m_{11} & m_{12} & \cdots & m_{1T} \\
 m_{21} & m_{22} & \cdots & m_{2T} \\
 \vdots & \vdots & \ddots & \vdots \\
 m_{N1} & m_{N2} & \cdots & m_{NT}
\end{pmatrix}_{N \times T},
\end{equation}

where $m_{ij}$, for $i=1,..., N$ and $j = 1,..., T$, is the cumulative variable of the $i$-th-node at time step $j$ (e.g., $P(X_i=1)$ in the SIS case or $\Phi_{i}$ for the PDG case). For convenient characterization, we represent the matrix $\mathcal{M}$ as $\mathcal{M}=(\mathcal{M}_{1}, \mathcal{M}_{2},..,\mathcal{M}_{t},...,\mathcal{M}_{T})$, where $\mathcal{M}_{t}$ is a column vector as defined before. Next, we encode the inner-interactions of node pairs $e(i, j)$, with $i,j\in\mathcal{V}$, at a time step $t$ from the marginal data $\mathcal{M}_{t}$ as $\delta^{t}$:

\begin{equation}
\label{eq:inner_interactions}
\delta^{t}_{i,j}=|m_{it}-m_{jt}|.
\end{equation}

Thus, the message difference for all $2N$ node pairs is represented as:

\begin{equation}
\label{eq:message_difference}
\Delta_t=
\begin{pmatrix}
 \delta^{t}_{1, 1} & \delta^{t}_{1, 2} & \cdots & \delta^{t}_{1, N} \\
 \delta^{t}_{2, 1} & \delta^{t}_{2, 2} & \cdots & \delta^{t}_{2, N} \\
 \vdots & \vdots & \ddots & \vdots \\
 \delta^{t}_{N, 1} & \delta^{t}_{N, 2} & \cdots & \delta^{t}_{N, N}
\end{pmatrix}_{N \times N}.
\end{equation}

We measure the message gap between pairs of nodes by the absolute value of the deviation of the marginal data. In practice, the definition of the message gap can be chosen arbitrarily; here, we have adopted the simplest form in Eq. \ref{eq:inner_interactions}. The effect of using different forms of deviation is not discussed here. In experiments, we prefer the squared deviation $(m_{it}-m_{jt})^2$ over the absolute one, as it has the advantage of magnifying the data differences while avoiding mathematical complexity. Further, we define the mapping $\xi(\cdot)$ in the form of an augmented matrix:

\begin{subequations} 
    \label{eq:augmented_matrix}
    \begin{align}
    \xi(\mathcal{M}_t) &= (A \odot \Delta_t , \mathcal{M}_t), \quad t=1,\cdots, T \label{eq:10A}\\
                    &=
                    \begin{pmatrix}
                     a_{11}\cdot\delta^{t}_{1, 1} & a_{12}\cdot\delta^{t}_{1, 2} & \cdots & a_{1N}\cdot\delta^{t}_{1, N} & m_{1t}\\
                     a_{21}\cdot\delta^{t}_{2, 1} & a_{22}\cdot\delta^{t}_{2, 2} & \cdots & a_{2N}\cdot\delta^{t}_{2, N} & m_{2t}\\
                     \vdots & \vdots & \ddots & \vdots \\
                     a_{N1}\cdot\delta^{t}_{N, 1} & a_{N2}\cdot\delta^{t}_{N, 2} & \cdots & a_{NN}\cdot\delta^{t}_{N, N} & m_{Nt}
                    \end{pmatrix}, \label{eq:10B}
    \end{align}
\end{subequations}

where $A \odot \Delta_t$ denotes the Hadamard product of matrices $A$ and $\Delta_t$, and $A$ represents the adjacency matrix encoding the system's network topology $\mathcal{G}(\mathcal{V},\mathcal{E})$:
\begin{equation}
    \label{eq:adjacency_matrix}
    A=
    \begin{cases}
    a_{ij} = 1,\quad e(i, j) \in \mathcal{E}, \\
    a_{ij} = 0,\quad otherwise.
    \end{cases}
\end{equation}

Through TMIFE, we obtain the encoding $\xi(\cdot)$, which provides features incorporating dynamic system interaction information to the message-passing based GCN model for collective behavior evolution prediction tasks.

\subsection{High-dimensional Behavior Prediction}
\label{subsec:prediction}

We consider only the case of undirected graphs $\mathcal{G}(\mathcal{V},\mathcal{E})$ in our work, with its adjacency matrix denoted as $A$. Since the information from the nodes themselves needs to be integrated into the messaging process, it is necessary to add a self-loop term to each node, i.e., $\tilde{A} = A + I_{N}$, where $I_N$ is an identity matrix of order $N$. The degree diagonal matrix is defined as $D = \text{diag}\left(\sum_{j=1}^{N}{A_{ij}}\right)$ for $i = 1, 2, \cdots, N$. The symmetrically normalized Laplacian matrix of $A$ is then given by:

\begin{equation}
\label{eq:laplacian}
L=I_N-(D^{+})^{\frac{1}{2}}A(D^{+})^{\frac{1}{2}},
\end{equation}
where $D^{+}$ is the Moore-Penrose inverse. Since the degree matrix $D$ is diagonal, its reciprocal square root $(D^{+})^{1/2}$ is just the diagonal matrix whose diagonal entries are the reciprocals of the square roots of the diagonal entries of $D$, thus we can express it as $(D^{+})^{1/2}=D^{-1/2}$. After applying the renormalization trick, it becomes: $L=I_N-D^{-\frac{1}{2}}AD^{-\frac{1}{2}} \longrightarrow \tilde{L}=\tilde{D}^{-\frac{1}{2}}\tilde{A}\tilde{D}^{-\frac{1}{2}}$, where $\tilde{D}_{ii}=\tilde{A}_{ii}$.

For a signal $\mathcal{X} \in \mathbb{R}^{N}$, the graph convolution can be expressed as a filter $g(\theta) = \text{diag}(\theta)$ parameterized by $\theta \in \mathbb{R}^{N}$ in the Fourier domain (Eq. \ref{eq:13A}), where $U$ is the matrix of eigenvectors obtained from the decomposition $L = U \Sigma U^{\prime}$. Therefore, $g(\theta)$ is a function of the eigenvalues of $L$, i.e., $g(\theta) = g_{\theta}(\Sigma)$. Using a first-order approximation of the Chebyshev polynomials, we have:

\begin{subequations} 
    \label{eq:chebishev}
        \begin{align}
        g(\theta)\mathcal{X} &= Ug(\theta)U^{\prime}, \label{eq:13A} \\
        g(\theta^{\prime})\mathcal{X} &\approx \theta^{\prime} (I_N+D^{-\frac{1}{2}}AD^{-\frac{1}{2}})\mathcal{X}, \label{eq:13B}
        \end{align}
\end{subequations}

where $\theta^{\prime} \in \mathbb{R}^{k}$ is a vector of Chebyshev coefficients, with $k = 1$. Following the signal filtering in the Fourier domain and the first-order approximation of the Chebyshev polynomials, we define a linear function on the graph Laplacian spectrum (graph convolution) as:

\begin{equation}
    \label{eq14}
    H^{(l+1)} = \sigma\left(\tilde{D}^{-\frac{1}{2}}\tilde{A}\tilde{D}^{-\frac{1}{2}}H^{(l)}W^{(l)}\right)
\end{equation}

where $W^{(l)}$ is a trainable weight matrix for the $l$-th layer, and $H^{(0)} = \mathcal{X}$.

We perform semi-supervised classification learning on the final graph embedding vectors using a Fully Connected Neural Network (FCNN) to obtain predictions of a node's state at any time $t$. The task can be expressed as:

\begin{equation}
\label{eq15}
Z = \sigma_{1}\left(\sigma_{2}\left(\tilde{D}^{-\frac{1}{2}}\tilde{A}\tilde{D}^{-\frac{1}{2}}H^{(l)}W^{(l)}\right)W^{(fc)}\right)
\end{equation}

where $\sigma_1$ is the ReLU activation function, and $\sigma_2$ is the Softmax activation function. It is important to note that we use TMIFE to encode $H_t^{(0)}$, i.e., $\mathcal{X} = \xi(\mathcal{M}_t)$ from Eq. \ref{eq:augmented_matrix}.

To perform the semi-supervised node classification task, we use the simple and efficient cross-entropy loss function $\mathcal{L}$:

\begin{equation}
    \label{eq:loss_function}
    \mathcal{L}(Y, Z) = -\sum_{y \in \Omega_\mathcal{Y}} \sum_{p=1}^{P}{\mathcal{Y}_{y,p} \textbf{ln}Z_{y,p}}
\end{equation}

where $P$ is the dimension of the graph embedding, and $\Omega_{\mathcal{Y}}$ is the set of nodes with labels. The algorithmic framework and pseudo-code for Sections \ref{subsec:tmife} and \ref{subsec:prediction} are illustrated in Algorithm \ref{alg:tmife_prediction}.

\begin{algorithm}[htbp]
\caption{Prediction with TMIFE in $t$-th time step.}\label{alg:alg1}
\begin{algorithmic}
\State 
\State {\textsc{TMIFE PROCESS}}
\State \hspace{0.15cm}$ \textbf{Initialize} \quad N, \quad T, \quad \mathbf{S}, \quad \mathcal{G}(\mathcal{V},\mathcal{E})$
\State \hspace{0.15cm}$ s^\mathbf{t}_\mathbf{i} \gets p\left(s^t_i \in S\right)$ \textbf{for} $ i= 1,\cdots,N $
\State \hspace{0.15cm}$ \mathbf{A} \gets  \{a_{ij}=1|e(i,j)\in \mathcal{E}\}$ \textbf{for} $i,j= 1,\cdots,N $
\State \hspace{0.15cm} \textbf{repeat} $\quad t$
\State \hspace{0.9cm}$ \mathcal{M}_\mathbf{t} \gets \{m_{it}=\mathcal{Q}\left(i, j|(s^t_i, s^t_j) \in \mathbf{S}\right)\} $ \textbf{for} $ e(i,j) \in \mathcal{E}$
\State \hspace{0.9cm}$ \Delta_t \gets \{\delta^t_{i,j}=|m_{it}-m_{jt}|\} $ \textbf{for} $i,j= 1,\cdots,N $
\State \hspace{0.9cm}$ \xi(\mathcal{M}_t) \gets \left(\mathbf{A}\odot \Delta_t, \mathcal{M}_t\right) $
\State \hspace{0.15cm} \textbf{until} $\quad t=T$
\State \hspace{0.15cm}\textbf{return}  $\quad \xi(\mathcal{M}_t),  \mathbf{S}$
\State 
\State {\textsc{TRAIN MODEL}}
\State \hspace{0.15cm}$ \textbf{Initialize} \quad \textbf{W}_{g}, \quad \textbf{W}_{m}$
\State \hspace{0.15cm}$ \mathcal{X}_\mathbf{t} \gets \xi\left(\mathcal{M}_t\right) $
\State \hspace{0.15cm}$ \mathbf{\tilde{A}}(\tilde{a}) \gets \mathbf{A}+=\mathbf{I_N} $
\State \hspace{0.15cm}$ \mathbf{\tilde{D}} \gets \{\tilde{d}_{ii}=\sum_{j\in \Omega_i}\tilde{a}_{ij}\}$ \textbf{for} $i,j= 1,\cdots,N $
\State \hspace{0.15cm}$ \mathbf{H^{(0)}_t} \gets \mathcal{X}_\mathbf{t} $
\State \hspace{0.15cm} \textbf{repeat} \quad \textbf{Epoch}
\State \hspace{0.9cm}$ \mathbf{\tilde{L}} \gets \mathbf{\tilde{D}}^{-1/2} \mathbf{\tilde{A}} \mathbf{\tilde{D}}^{-1/2}$
\State \hspace{0.9cm}$ \mathbf{H^{l+1}}_\mathbf{t} \gets ReLU(\left(\mathbf{\tilde{L}}\mathbf{H^{(l)}_t}\mathbf{W_g^{(l)}}+\mathbf{B_g}\right) $ \textbf{for} $l=1,2$
\State \hspace{0.9cm}$\mathbf{Z_t} \gets Softmax\left(\mathbf{H^{(2)}_t}\mathbf{W_m}+\mathbf{B_m}\right) $ \textbf{for} $l=1,2$
\State \hspace{0.9cm} \textbf{Loss} $\gets \mathcal{L}(\mathbf{Z_t}, S_t), \quad S_t \in \mathbf{S}$
\State \hspace{0.9cm} \textbf{backward propagation with Loss}
\State \hspace{0.9cm} \textbf{update}  $\mathbf{W_g},\mathbf{W_m}$
\State \hspace{0.15cm} \textbf{until} \quad $\textbf{end Epoch}$
\State \hspace{0.15cm} \textbf{return} $\mathbf{Z_t}, \mathbf{H^{(2)}_t}$
\end{algorithmic}
\label{alg:tmife_prediction}
\end{algorithm}

\section{Numerical Simulation Settings}
\label{sec:numerical}

In this section, we explain the explain the experimental settings where the prediction tasks are carried. The purpose of these experiments is to verify that our proposed TMIFE method can accurately predict the state of nodes with the assistance of GCNs. More importantly, the prediction process should also have the capability to generalize across different time scales and make accurate predictions about the evolutionary state of inner-interactive topological systems at arbitrary time steps.

\begin{itemize} 
\item{\textbf{Networks}} All the dynamical process studied in this work run on some type of network topology. We adopt three diverse and well-known network models, $\mathcal{G}(\mathcal{V},\mathcal{E})$ with size $N$ and $m$ edges: the Square Lattice (SL) network with periodic boundary conditions, the Barabási-Albert (BA) scale-free network with number of links parameter $m_0$, and the Erdős-Rényi (ER) random network with connection probability $p$. We control the average degree $\langle k \rangle$ by reparameterizing the BA and ER networks, while the SL network has a constant von Neumann neighborhood of $\langle k\rangle=4$. For the ER network, $\langle k \rangle = p(N-1)$; for the BA network, $\langle k \rangle = 2m_0$. These topologies results may in dramatically different dynamics for the dynamics processes run on them, thus presenting a challenge for node-status prediction. We conducted a wide range of experiments on the SL, BA, and ER networks in various scenarios as follows.

\item{\textbf{Collective Behavior Dynamics Prediction}} In the dual-strategy PDG, we perform three levels of prediction: (1) Predict the behavioral evolutionary trajectories of all individuals in the remaining rounds based on the marginal data from the first $r$ rounds. (2) Predict the phase transition of the cooperation fraction $f_C$ in the evolutionary steady state over the parameter space of the temptation parameter $b$, and determine the prediction threshold $\mu_0$. (3) Predict the spatiotemporal evolution pattern of collective behavior on the SL network.

\hspace{0.2cm} To further clarify, the training set consists of $\mathcal{M}_t$ for the first $r$ time steps and the adjacency matrix $A$, corresponding to the labels $S_t=\left(s_1^t,\cdots,s_n^t \right), n<N$. $\mathcal{M}_t$ needs to be obtained using the proposed TMIFE method, and the test set is used to evaluate the model's performance. We then directly predict the states of all nodes for the remaining time steps using the trained model and assess the prediction accuracy of the evolutionary curves. Additionally, we extend the dual-strategy model to a $3$-strategy extension of the PDG (described before in \ref{subsec:games}), which includes the exit strategy $E$. The modeling process and objectives remain the same as in the $2$-strategy case.

\item{\textbf{Real Human Behavior Experiments}} In social sciences and economics, behavioral experiments are critical for understanding human decision-making processes. We designed offline, real human repeated Prisoner's Dilemma experiments and collected behavioral data on human decision-making. Specific experimental details are provided here in section \ref{sec:human}. We adapted our model to the PDG experimental data collected above. The experiment uses a $7 \times 7$ SL network with periodic boundary conditions, typical of anonymous symmetrical game experiments. Forty-nine individuals participated, and the evolution continued for a total of $50$ rounds. We trained the model on the first $3$ rounds and predicted the evolution curve $f_C$ for all $50$ rounds. It is important to note that this experiment recorded the details of each individual's strategy in every round of the pairwise game, allowing us to compute cumulative payoff data for the marginal information matrix $\mathcal{M}_t$ at any time step.

\item{\textbf{SIS Dynamics}} To investigate whether the proposed TMIFE method is capable of generalization across different tasks, we applied it to a dynamical process other than evolutionary games, such as the well-known susceptible-infected-susceptible (SIS) for epidemic spreading. The dynamics and information propagation mechanism of the SIS is entirely different from that of PDG, where the infection probability of the focal node is determined by the number of infected neighbors. 

\item{\textbf{Migration Learning Under Different Social Dilemmas}} We also tested the model under more challenging conditions by training it in the PDG scenario and performing zero-shot inference in the SDG, HG, and SHG scenarios. Our aim is to test the generalization ability of our model and confirm whether it can truly learn the rules of collective behavior decision-making using only marginal payoff data. Transferability is crucial for designing agents that are robust in diverse environments, so we require the model to correctly identify human decision-making rules under different social dilemmas, even if it has been trained only once in a specific scenario. \end{itemize}

\section{Baselines}
\label{sec:baselines}

To evaluate performance, we applied various supervised and unsupervised methods and compared them with our proposed TMIFE method. To assess the model's training effectiveness and generalization capability, we tested the model's performance on two tasks: (1) accuracy for the node-status classification task on the test dataset, and (2) accuracy for the evolutionary curve on the zero-shot dataset. Additionally, we conducted ablation experiments on the TMIFE method.

\begin{itemize} \item{\textbf{Node2Vec}} Node2Vec is a simple and effective unsupervised graph embedding algorithm that encodes higher-order information of nodes in the graph through biased random walks. We control the balance between depth-first search (DFS) and breadth-first search (BFS) by tuning the exploration parameters $p$ and $q$, adjusting the model to optimal conditions. Since Node2Vec is a transductive method, meaning it can only embed nodes it has already seen during training, we use parts of the test set to compare its performance with inductive methods, which can generalize to unseen nodes.

\item{\textbf{G}raph \textbf{Sa}mple and A\textbf{g}gregat\textbf{e} (\textbf{GraphSAGE})} addresses the efficiency problem in scenarios with a large number of nodes by reducing the complexity of aggregating all neighbors' information through a sampling process. This extends GCN to the inductive learning framework. We use a supervised form of GraphSAGE for graph representation learning and apply it to a downstream node classification task.

\item{\textbf{G}raph \textbf{At}tention \textbf{N}etwork (\textbf{GAT})} considers the importance of each node and its neighbors, prioritizing the aggregation of information within the neighborhood. GAT has achieved good results in both inductive and transductive learning tasks. We perform a supervised node prediction task using GAT and optimize its parameters through grid search.

\item{\textbf{G}raph \textbf{I}somorphism \textbf{N}etwork 
(\textbf{GIN})} aims to improve the expressive power of graph neural networks to capture graph structures accurately. The key idea is that two isomorphic (structurally identical) graphs should have the same representation, while non-isomorphic graphs should have different representations. GIN achieves this by using an aggregation function that is injective (one-to-one), such as a summation function, which is more expressive than averaging or mean-pooling used in other GCNs. After aggregating the features of the neighbors, GIN combines them with the node's original features and processes them using a fully connected network that can approximate any function, thereby enhancing its representation power. We apply GIN in a supervised learning context to perform node classification and evolution prediction tasks.
\end{itemize}

\section{Parameter Settings}
\label{sec:parameters}

We select the initial $r$ rounds from the total $T$ rounds of dynamics to form the training and testing sets, while the remaining $T - r$ rounds are used as the zero-shot dataset to evaluate generalization ability. The first $r$ rounds of data are divided into batches of a fixed size, with a \textbf{batch size} of 8 and $(r/T) \times 100 = 20\%$. Note that this batching refers to the segmentation of data across different time steps, not the division of the training and testing sets. This is because GCN requires the full network structure and cannot perform semi-supervised transductive learning by dividing the batch.

The proportions of the training set, validation set, and test set are $\{75\%,3\%,22\%\}$. We find the optimal parameters by tuning them on the validation set using grid search. During the gradient descent process, we utilize the Adam optimizer with a decay rate of $\lambda=1\times 10^{-5}$ and determine the optimal learning rate $\alpha$ through grid search. During training, we apply L2 regularization to the loss function to reduce the risk of overfitting. We also add small Gaussian noise to the marginal data to improve training set continuity.

Regarding population size, we consistently use $N=50\times 50$ and control the ER and BA networks by varying the average degree $\langle k\rangle= \{2,4,6,8,10\}$. The noise parameter involved in the Fermi updating rule (Eq. \ref{eq:fermi_updating}) is set to $\kappa=0.1$. The infection and recovery rates mediating, respectively, the defined related of the main text are $\beta=\{0.01,0.3, 0.7\}$ and $\gamma=0.2$, respectively. The number of convolutional layers is $l=2$, the graph embedding output dimension is $\text{dim}(H^{l})=32$, and the output dimension of the downstream fully connected layer is $2$. For the three-strategy extended PDG case, the output dimension of the fully connected layer is $3$.

The equipment used to train the model includes an Intel i5 13600k CPU and an NVIDIA RTX 4060ti 16GB GPU. For reference, training a prediction model in a PDG scenario on a $50 \times 50$ network takes approximately $0.7$ GPU months.

\section{Evaluation Metrics}
\label{sec:evaluation}

We use various metrics to evaluate the model's performance. Accuracy measures global accuracy, while the F1-score is suitable for imbalanced samples. For node classification on the test set and the zero-shot set, we use Accuracy (\textbf{Acc}), F1-score (\textbf{F1}), True Positive Rate (\textbf{TPR}), and False Positive Rate (\textbf{FPR}) to assess effectiveness. For the evolutionary curve prediction on the zero-shot set, we use Mean Squared Error (\textbf{MSE}), Mean Absolute Error (\textbf{MAE}), and Jensen-Shannon divergence (\textbf{JS}) to measure the approximation between the predicted and true curves.

\textbf{Acc}uracy is a common metric for classification tasks that measures the correct rate and is known for its convenience and generalizability:

\begin{equation}
    \label{eq17}
    Acc = \frac{TP+TN}{TP+TN+FP+FN},
\end{equation}

where \textbf{TP} (True Positive), \textbf{FP} (False Positive), \textbf{TN} (True Negative), and \textbf{FN} (False Negative). The F1-score (\textbf{F1}), \textbf{TPR}, and \textbf{FPR} are derived from the confusion matrix of the classification task. The \textbf{F1} score combines information from both positive and negative samples, performing well with imbalanced data. \textbf{TPR} and \textbf{FPR} assess classification accuracy from the perspectives of the true positive rate and the false positive rate, respectively:

\begin{subequations} 
    \label{eq:18}
        \begin{align}
        F1 &= \frac{2\times Precision \times Recall}    {Precision+Recall}, \label{eq:18A} \\
        TPR &= Recall = \frac{TP}{TP+FN}, \label{eq:18B} \\
        FPR &= \frac{FP}{TN+FP}, \label{eq:18C}
    \end{align}
\end{subequations}

where \textbf{Precision}=$\frac{\text{TP}}{\text{TP}+\text{FP}}$. In the $3$-strategy case, only global accuracy is considered.

\textbf{MSE} and \textbf{MAE} compare the differences between the real and predicted evolution curves on a point-by-point basis:

\begin{subequations} 
    \label{eq:19}
    \begin{align}
        MAE &= \frac{1}{T} \sum_{t=1}^{T}{\left| y_t-\hat{y}_t \right|}, \label{eq:19A} \\
        MSE &= \frac{1}{T} \sum_{t=1}^{T}{\left( y_t-\hat{y}_t \right)^2}. \label{eq:19B}
    \end{align}
\end{subequations}

where $\hat{y}_t$ is the average of the predictions over multiple runs ($100$ runs). For robustness, Jensen-Shannon divergence (\textbf{JS}) can be used to measure the similarity of curve predictions from a distributional perspective:

\begin{subequations} 
    \label{eq:20}
    \begin{align}
        KL\left(P||Q\right) &=  \sum_{T} P(y_t)\frac{P(y_t)}{Q(\hat{y}_t)}\label{eq:20A}, \\
        JS\left(P_y||P_{\hat{y}}\right) &= \frac{1}{2}KL\left(P_y||\frac{P_y+P_{\hat{y}}}{2}\right) + \frac{1}{2}KL\left(P_{\hat{y}}||\frac{P_y+P_{\hat{y}}}{2}\right). \label{eq:20B}
    \end{align}
\end{subequations}

\section{Real-Human Prisoner's Dilemma Game Experimental Details}
\label{sec:human}

This appendix presents the time, location, participants, experimental setup, and the results and data composition of the real-person experiment we organized.

The experiment was conducted at Yunnan University of Finance and Economics on March 10, 2018, at 9:30 a.m. Beijing time and was organized by Prof. Lei Shi's team.

There were $49$ participants in this experiment (recruited voluntarily), all from Yunnan University of Finance and Economics, randomly sampled from first- to third-year undergraduate students. The experiment was set up on a $7 \times 7$ Square Lattice network with periodic boundary conditions. Participants were randomly numbered by a computer and randomly matched, so no one knew who they were playing against, ensuring the experiment remained anonymous. Each subject could choose either cooperation ($C$) or defection ($D$) as a strategy to play in the pairwise games. The benefits that individuals could obtain were determined by the following payoff matrix:

\begin{equation}
    \label{eq1}
    \begin{array}{l}
    \;\;\;\;\;\;\;\;\;\;\;
    \begin{array}{*{20}c}
        {C} & {D}  \\
    \end{array} \\ 
    \begin{array}{*{20}c}
        C  \\
        D  \\
    \end{array}\;
        \left( {
        \begin{array}{*{20}c}
            2 & -2  \\
            4 & 0 \\
        \end{array}
        } \right),   \\ 
    \end{array}
\end{equation}

\begin{figure}[htbp] 
    \centering
    \includegraphics[width=1\linewidth]{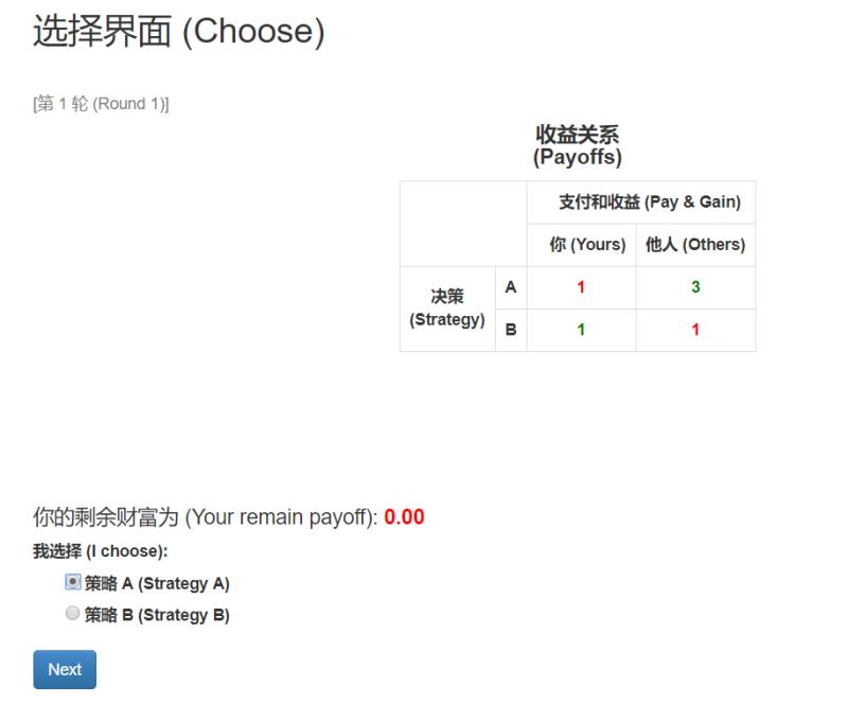} \caption{Experimental interface display 1.} 
    \label{fig:human_display1} 
\end{figure}

\begin{figure}[htbp] 
    \centering 
    \includegraphics[width=1\linewidth]{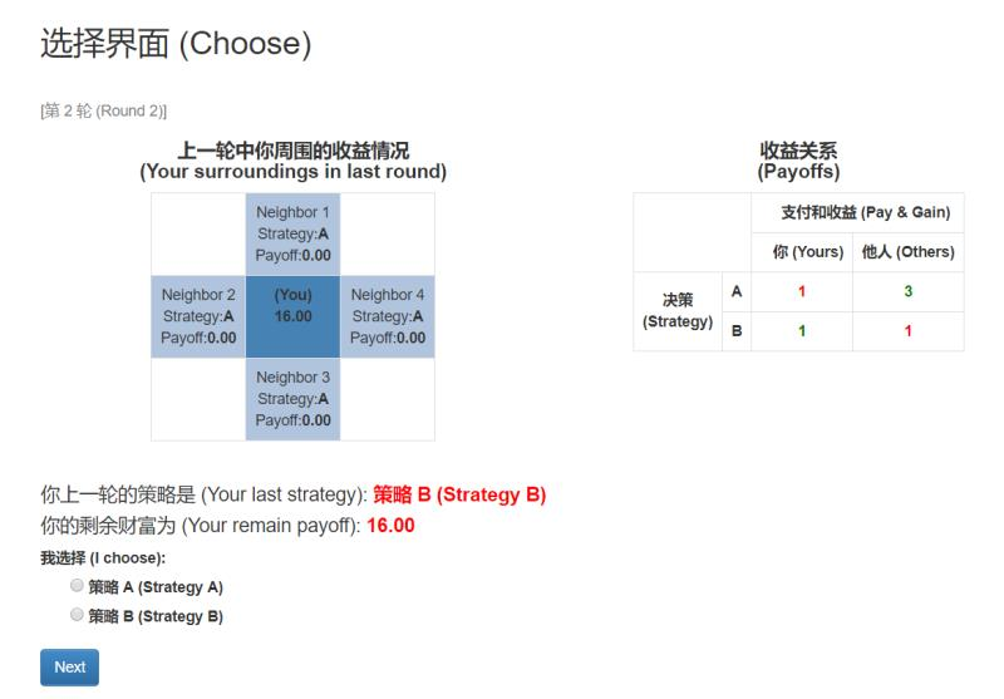} \caption{Experimental interface display 2.} 
    \label{fig:human_display2} 
\end{figure}

Participants adopted a ``point strategy,'' meaning the strategy chosen by each individual remains fixed for all of their interactions with their neighbors for a whole round. The experiment lasted a total of $50$ rounds. At the end of the experiment, the actual gain (in RMB) for an individual was calculated as: $15$ (appearance fee) + (final score) $\times 0.2$. Subjects made their strategy choices through an interface displayed on a computer screen. We recorded two types of experimental information: (1) the subjects' cumulative payoffs per round (marginal data) and (2) the subjects' strategies per round (labels), in terms of focal individuals. The experimental interface is shown in Figures \ref{fig:human_display1} and \ref{fig:human_display2}.

\section{Supplementary Results}
\label{sec:results}

\begin{table*}[htbp]
\centering
\caption{Test set evaluation for dual-strategy PDG. \label{tab:table1}}
\begin{tabular}{ccccccccc}
\hline
\multicolumn{1}{l}{}       & \multicolumn{1}{l}{Acc} & \multicolumn{1}{l}{F1} & \multicolumn{1}{l}{TPR} & \multicolumn{1}{l}{FPR}     & \multicolumn{1}{l}{Acc} & \multicolumn{1}{l}{F1} & \multicolumn{1}{l}{TPR} & \multicolumn{1}{l}{FPR} \\ \hline
$\langle k \rangle$ & \multicolumn{4}{c|}{BA}                                                                                  & \multicolumn{4}{c}{BA std.}                                                                          \\ \hline
2                          & 0.9748                  & 0.9815                 & 0.9828                  & \multicolumn{1}{c|}{0.3034} & 0.0473                  & 0.0444                 & 0.0554                  & 0.2771                  \\
4                          & 0.9548                  & 0.9721                 & 0.9543                  & \multicolumn{1}{c|}{0.0158} & 0.1067                  & 0.0692                 & 0.1083                  & 0.0805                  \\
6                          & 0.9087                  & 0.9499                 & 0.9187                  & \multicolumn{1}{c|}{0.4592} & 0.1469                  & 0.0937                 & 0.1469                  & 0.2919                  \\
8                          & 0.7217                  & 0.8281                 & 0.7217                  & \multicolumn{1}{c|}{0.6540} & 0.1504                  & 0.1126                 & 0.1504                  & 0.1743                  \\
10                         & 0.4993                  & 0.4878                 & 0.7840                  & \multicolumn{1}{c|}{0.7002} & 0.0556                  & 0.1573                 & 0.1251                  & 0.1509                  \\ \hline
                           & \multicolumn{4}{c|}{ER}                                                                                  & \multicolumn{4}{c}{ER std.}                                                                          \\ \hline
2                          & 0.8623                  & 0.9004                 & 0.9278                  & \multicolumn{1}{c|}{0.3581} & 0.0796                  & 0.0788                 & 0.1244                  & 0.0719                  \\
4                          & 0.8757                  & 0.9087                 & 0.9026                  & \multicolumn{1}{c|}{0.7783} & 0.1255                  & 0.1142                 & 0.1474                  & 0.2799                  \\
6                          & 0.8731                  & 0.9224                 & 0.8743                  & \multicolumn{1}{c|}{0.1348} & 0.1545                  & 0.0982                 & 0.1522                  & 0.2854                  \\
8                          & 0.7552                  & 0.8480                 & 0.7773                  & \multicolumn{1}{c|}{0.6899} & 0.1151                  & 0.0912                 & 0.1259                  & 0.1959                  \\
10                         & 0.7582                  & 0.8483                 & 0.7653                  & \multicolumn{1}{c|}{0.7013} & 0.1626                  & 0.1148                 & 0.1686                  & 0.2954                  \\ \hline
                           & \multicolumn{4}{c|}{SL}                                                                                  & \multicolumn{4}{c}{SL std.}                                                                          \\ \hline
4                          & 0.7607                  & 0.6854                 & 0.7875                  & \multicolumn{1}{c|}{0.2622} & 0.0543                  & 0.1275                 & 0.0724                  & 0.0620                  \\ \hline
\end{tabular}
\end{table*}

\begin{table}[htbp]
\centering
\caption{Zero-shot evolutionary curve prediction for dual-strategy PDG. \label{tab:table2}}
\begin{tabular}{cccclll}
\hline
\multicolumn{1}{l}{}       & \multicolumn{1}{l}{MSE}    & \multicolumn{1}{l}{MAE}    & \multicolumn{1}{l}{JS}      & MSE     & MAE     & JS      \\ \hline
$\langle k \rangle$ & \multicolumn{3}{c|}{BA}                                                               & \multicolumn{3}{c}{BA std.} \\ \hline
2                          & 0.0004                     & 0.0090                     & \multicolumn{1}{c|}{0.0082} & 0.0011  & 0.0075  & 0.2670  \\
4                          & 0.0009                     & 0.0183                     & \multicolumn{1}{c|}{0.0143} & 0.0051  & 0.0073  & 0.2814  \\
6                          & 0.0033                     & 0.0352                     & \multicolumn{1}{c|}{0.0504} & 0.0095  & 0.0138  & 0.3047  \\
8                          & 0.0065                     & 0.0573                     & \multicolumn{1}{c|}{0.1450} & 0.0121  & 0.0190  & 0.3189  \\
10                         & 0.0072                     & 0.0269                     & \multicolumn{1}{c|}{0.0421} & 0.0177  & 0.0732  & 0.2666  \\ \hline
                           & \multicolumn{3}{c|}{ER}                                                               & \multicolumn{3}{c}{ER std.} \\ \hline
2                          & 0.0009                     & 0.0230                     & \multicolumn{1}{c|}{0.0183} & 0.0003  & 0.0093  & 0.0107  \\
4                          & 0.0014                     & 0.0345                     & \multicolumn{1}{c|}{0.0209} & 0.0007  & 0.0074  & 0.0308  \\
6                          & 0.0026                     & 0.0352                     & \multicolumn{1}{c|}{0.0603} & 0.0093  & 0.0038  & 0.0834  \\
8                          & 0.0029                     & 0.0304                     & \multicolumn{1}{c|}{0.0247} & 0.0468  & 0.0649  & 0.0801  \\
10                         & 0.0059                     & 0.0522                     & \multicolumn{1}{c|}{0.1508} & 0.0433  & 0.0506  & 0.1327  \\ \hline
                           & \multicolumn{3}{c|}{SL}                                                               & \multicolumn{3}{c}{SL std.} \\ \hline
4                          & \multicolumn{1}{l}{0.0016} & \multicolumn{1}{l}{0.0335} & \multicolumn{1}{l|}{0.0376} & 0.0007  & 0.0015  & 0.1517  \\ \hline
\end{tabular}
\end{table}

\begin{table*}[htbp]
\centering
\caption{Baseline comparison 1: Prediction of evolutionary behavior on the test set. \label{tab:table3}}
\begin{tabular}{lllllllll}
\hline
                & ACC    & F1     & TPR    & FPR    & ACC\_std & F1\_std & TPR\_std & FPR\_std \\ \hline
Node2Vec        & 0.4978 & 0.2780 & 0.2044 & 0.2114 & 0.0865   & 0.1387  & 0.1129   & 0.1176   \\
GAT             & 0.7022 & 0.4725 & 0.6752 & 0.2891 & 0.1378   & 0.1501  & 0.1881   & 0.1459   \\
GIN             & 0.7603 & 0.5745 & 0.5693 & 0.1499 & 0.0819   & 0.1542  & 0.1884   & 0.0910   \\
GCN             & 0.8033 & 0.6984 & 0.3302 & 0.1676 & 0.0940   & 0.1830  & 0.2830   & 0.0895   \\
GraphSAGE       & 0.6037 & 0.4485 & 0.7717 & 0.4315 & 0.0788   & 0.1656  & 0.1546   & 0.1018   \\ \hline
Node2Vec+TMIFE  & 0.5004 & 0.4098 & 0.3503 & 0.3405 & 0.0407   & 0.0731  & 0.0714   & 0.0831   \\
GAT+TMIFE       & 0.8627 & 0.8051 & 0.7307 & 0.0527 & 0.0751   & 0.1037  & 0.1199   & 0.0643   \\
GIN+TMIFE       & 0.8602 & 0.6951 & 0.6312 & 0.0575 & 0.0882   & 0.1585  & 0.1893   & 0.0589   \\
GCN+TMIFE       & \textbf{0.9442} & 0.8423 & \textbf{0.8297} & \textbf{0.0288} & 0.0997   & 0.3230  & 0.3125   & 0.0705   \\
GraphSAGE+TMIFE & 0.9002 & \textbf{0.8660} & 0.8076 & 0.0371 & 0.0818   & 0.0975  & 0.1187   & 0.0591   \\ \hline
\end{tabular}
\end{table*}

\begin{table}[htbp]
\centering
\caption{Baseline comparison 2: Prediction of zero-shot evolutionary curves. \label{tab:table4}}
\begin{tabular}{llll}
\hline
                & MSE    & MAE    & JS     \\ \hline
Node2Vec        & 0.0944 & 0.2906 & 3.7157 \\
GAT             & 0.0209 & 0.1212 & 1.0411 \\
GIN             & 0.0627 & 0.2405 & 3.2327 \\
GCN             & 0.0717 & 0.1276 & 1.8598 \\
GraphSAGE       & 0.0179 & 0.1251 & 1.0023 \\ \hline
Node2Vec+TMIFE  & 0.0385 & 0.1654 & 1.1183 \\
GAT+TMIFE       & 0.0078 & 0.0725 & 0.2387 \\
GIN+TMIFE       & 0.0061 & 0.0450 & 0.1733 \\
GCN+TMIFE       & \textbf{0.0016} & \textbf{0.0335} & \textbf{0.0376} \\
GraphSAGE+TMIFE & 0.0024 & 0.0379 & 0.0863 \\ \hline
\end{tabular}
\end{table}

\begin{table*}[htbp]
\centering
\caption{Test set classification task evaluation for SIS dynamics. \label{tab:table5}}
\begin{tabular}{ccccccccc}
\hline
\multicolumn{1}{l}{}       & \multicolumn{1}{l}{Acc} & \multicolumn{1}{l}{F1} & \multicolumn{1}{l}{TPR} & \multicolumn{1}{l}{FPR}     & \multicolumn{1}{l}{Acc} & \multicolumn{1}{l}{F1} & \multicolumn{1}{l}{TPR} & \multicolumn{1}{l}{FPR} \\ \hline
$\langle k \rangle$ & \multicolumn{4}{c|}{BA}                                                                                  & \multicolumn{4}{c}{BA std.}                                                                          \\ \hline
2                          & 0.9261                  & 0.8263                 & 0.8095                  & \multicolumn{1}{c|}{0.0502} & 0.0751                  & 0.0897                 & 0.1232                  & 0.1235                  \\
4                          & 0.7930                  & 0.7202                 & 0.6484                  & \multicolumn{1}{c|}{0.1013} & 0.0400                  & 0.0542                 & 0.0628                  & 0.0406                  \\
6                          & 0.8135                  & 0.6856                 & 0.6435                  & \multicolumn{1}{c|}{0.1025} & 0.0462                  & 0.0756                 & 0.0862                  & 0.0390                  \\
8                          & 0.8092                  & 0.6044                 & 0.6400                  & \multicolumn{1}{c|}{0.1314} & 0.0435                  & 0.1044                 & 0.1148                  & 0.0393                  \\
10                         & 0.7292                  & 0.5306                 & 0.4683                  & \multicolumn{1}{c|}{0.1491} & 0.0559                  & 0.0873                 & 0.0836                  & 0.0951                  \\ \hline
                           & \multicolumn{4}{c|}{ER}                                                                                  & \multicolumn{4}{c}{ER std.}                                                                          \\ \hline
2                          & 0.7270                  & 0.7287                 & 0.7111                  & \multicolumn{1}{c|}{0.2567} & 0.0491                  & 0.0388                 & 0.0507                  & 0.1113                  \\
4                          & 0.7771                  & 0.6552                 & 0.6484                  & \multicolumn{1}{c|}{0.1622} & 0.0524                  & 0.0592                 & 0.0708                  & 0.0891                  \\
6                          & 0.8027                  & 0.6087                 & 0.6291                  & \multicolumn{1}{c|}{0.1458} & 0.0415                  & 0.0807                 & 0.0855                  & 0.0943                  \\
8                          & 0.7953                  & 0.5739                 & 0.5768                  & \multicolumn{1}{c|}{0.1346} & 0.0371                  & 0.0863                 & 0.0935                  & 0.0470                  \\
10                         & 0.7376                  & 0.4340                 & 0.4328                  & \multicolumn{1}{c|}{0.1673} & 0.0470                  & 0.1171                 & 0.1144                  & 0.0386                  \\ \hline
                           & \multicolumn{4}{c|}{SL}                                                                                  & \multicolumn{4}{c}{SL std.}                                                                          \\ \hline
4                          & 0.8144                  & 0.7045                 & 0.6224                  & \multicolumn{1}{c|}{0.0802} & 0.0580                  & 0.0583                 & 0.0795                  & 0.0975                  \\ \hline
\end{tabular}
\end{table*}

\newpage

\begin{table}[htbp]
\centering
\caption{Zero-shot evolutionary curve prediction for SIS dynamics. \label{tab:table6}}
\begin{tabular}{cccc}
\hline
\multicolumn{1}{l}{}       & \multicolumn{1}{l}{MSE} & \multicolumn{1}{l}{MAE} & \multicolumn{1}{l}{JS} \\ \hline
$\langle k \rangle$ & \multicolumn{3}{c}{BA}                                                     \\ \hline
2                          & 0.0031                  & 0.0398                  & 0.0897                 \\
4                          & 0.0012                  & 0.0285                  & 0.0299                 \\
6                          & 0.0029                  & 0.0397                  & 0.0998                 \\
8                          & 0.0041                  & 0.0562                  & 0.1437                 \\
10                         & 0.0043                  & 0.0552                  & 0.1738                 \\ \hline
                           & \multicolumn{3}{c}{ER}                                                     \\ \hline
2                          & 0.0069                  & 0.0444                  & 0.1545                 \\
4                          & 0.0028                  & 0.0346                  & 0.0070                 \\
6                          & 0.0022                  & 0.0331                  & 0.0420                 \\
8                          & 0.0025                  & 0.0439                  & 0.0593                 \\
10                         & 0.0028                  & 0.0416                  & 0.0709                 \\ \hline
                           & \multicolumn{3}{c}{SL}                                                     \\ \hline
4                          & 0.0017                  & 0.0256                  & 0.0598                 \\ \hline
\end{tabular}
\end{table}

\begin{figure*}[htbp]
 \centering
 \includegraphics[width=1\linewidth]
 {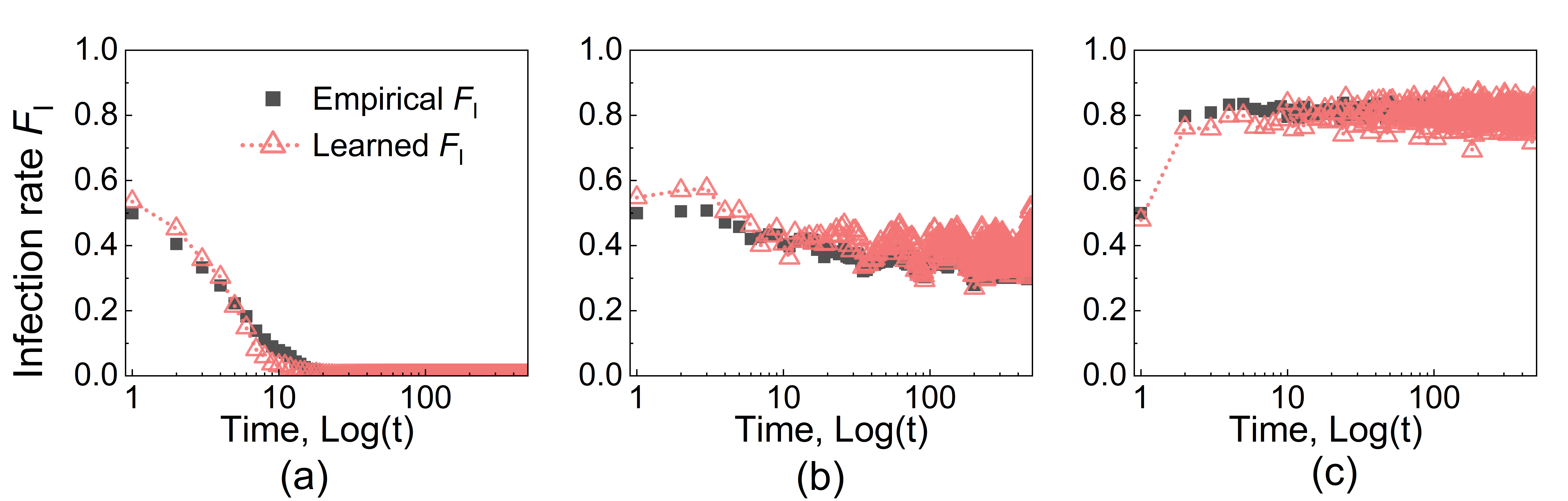}
 \caption{Prediction of SIS epidemic dynamics. Initial condition for nodes statuses is randomly assigned. Black squares and red triangles represent the simulation and predicted disease prevalence, respectively. Epidemiological parameters: recovery rate $\gamma=0.2$, and transmission rate $\beta=0.01$ (left panel), $\beta=0.3$ (center panel), and $\beta=0.7$ (right panel).}
 \label{fig:fig_sm_3_sis}
\end{figure*}

\begin{figure}[htbp]
 \centering
 \includegraphics[width=1\linewidth]
 {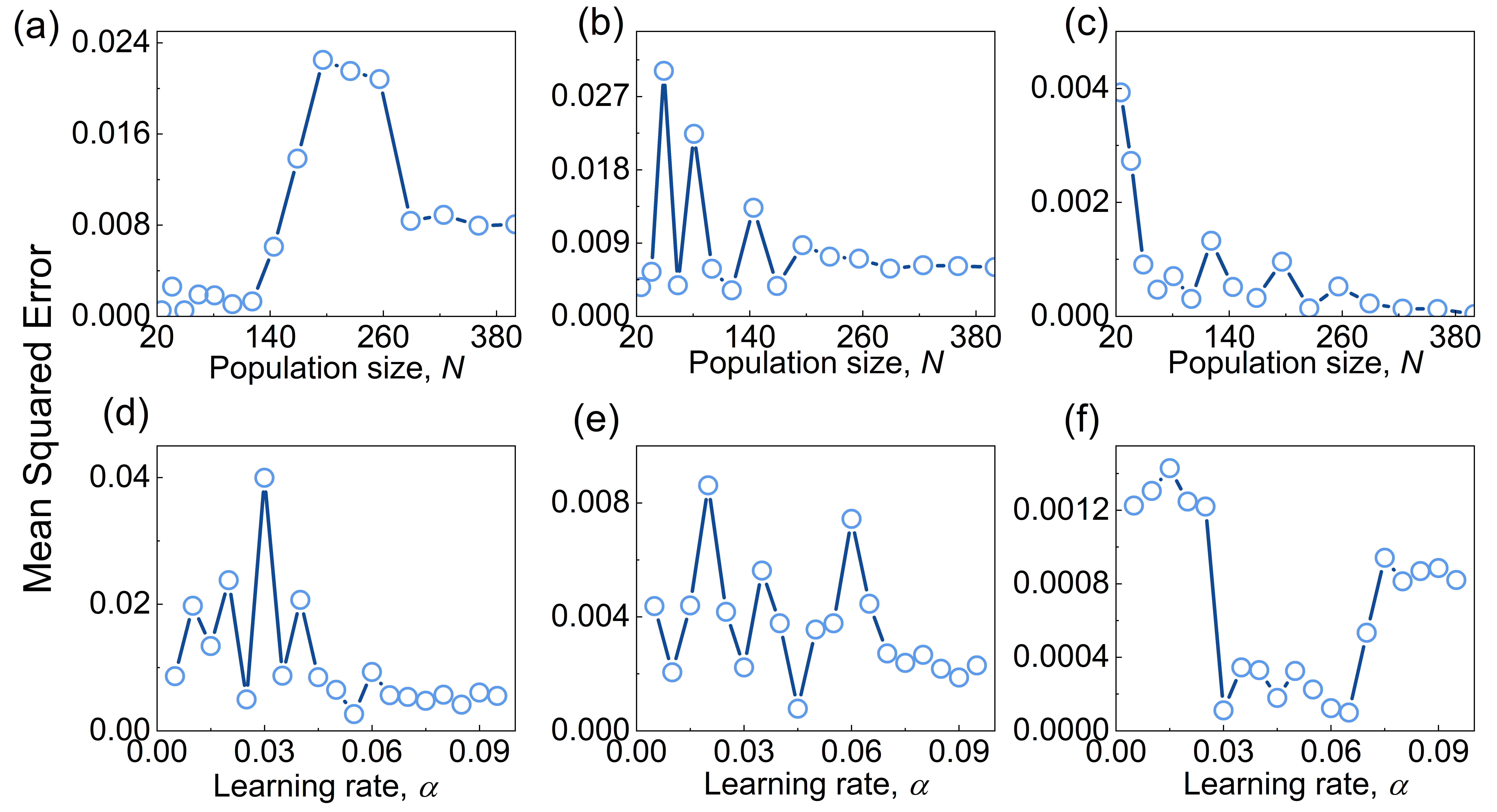}
 \caption{Robustness tests of the model with respect to the population size $N$ in panels (a) to (c), and to the learning rate $\alpha$ in panels (d) to (f).}
 \label{fig:fig_sm_4_robustness}
\end{figure}

\begin{figure}[htbp]
 \centering
 \includegraphics[width=1\linewidth]
 {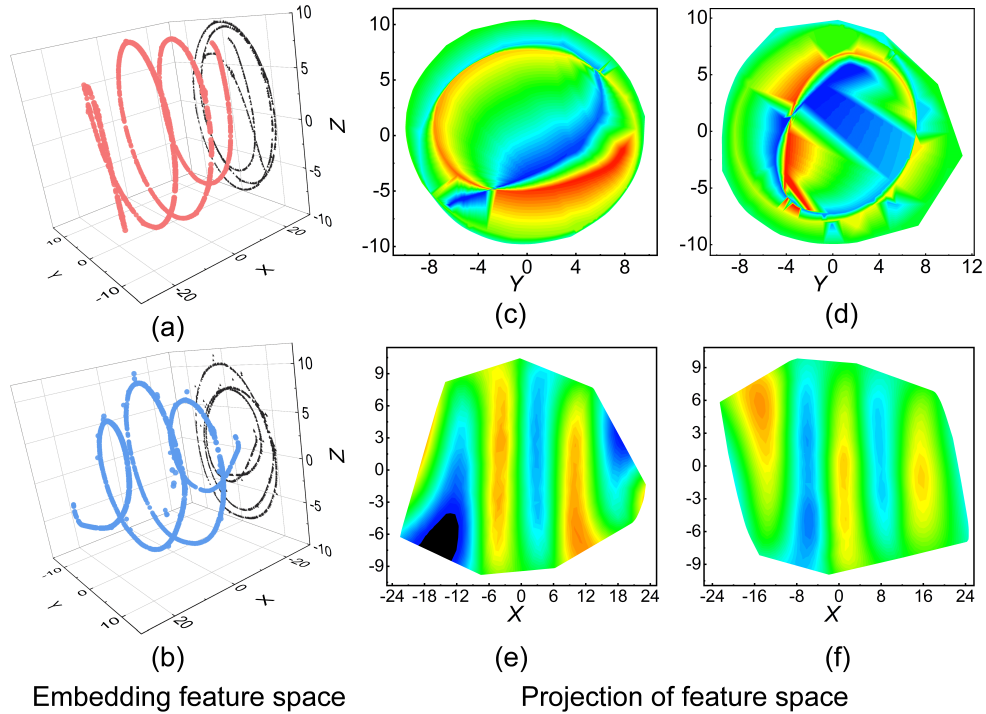}
 \caption{Visualization of the embedding space obtained by the TMIFE method. (a)-(b) represent $3$-dimensional TMIFE-GCN embedding vectors. The TMIFE (top panels) and the adjacency matrix (bottom panels) scenarios are compared, respectively. Panels (c) to (f) show the $2$-dimensional unfolding of the $3$-dimensional embedding space, where panels (c) and (e) are TMIFE cases and panels (d) and (f) are adjacency matrix cases.}
 \label{fig:fig_sm_5_embedding}
\end{figure}

\bibliography{sm_references}